\documentclass[%
 aip,
 amsmath,amssymb,
 reprint,%
]{revtex4-1}

\usepackage{graphicx}
\usepackage{dcolumn}
\usepackage{bm}

\usepackage[utf8]{inputenc}
\usepackage[T1]{fontenc}
\usepackage{mathptmx}
\usepackage{etoolbox}
\usepackage{xcolor}

\makeatletter
\def\@email#1#2{%
 \endgroup
 \patchcmd{\titleblock@produce}
  {\frontmatter@RRAPformat}
  {\frontmatter@RRAPformat{\produce@RRAP{*#1\href{mailto:#2}{#2}}}\frontmatter@RRAPformat}
  {}{}
}%
\makeatother
\begin{document}

\preprint{AIP/123-QED}

\title[Ab initio quasi-harmonic thermoelasticity of molybdenum at high temperature and pressure]{Ab initio quasi-harmonic thermoelasticity of molybdenum at high temperature and pressure}

\author{X. Gong}
 \affiliation{International School for Advanced Studies (SISSA), Via Bonomea 265, 34136, Trieste, Italy}
  \affiliation{IOM - CNR, Via Bonomea 265, 34136, Trieste, Italy}
 \email{xgong@sissa.it}

\author{A. Dal Corso}
 \affiliation{International School for Advanced Studies (SISSA), Via Bonomea 265, 34136, Trieste, Italy}
 \affiliation{IOM - CNR, Via Bonomea 265, 34136, Trieste, Italy}

\date{\today}

\begin{abstract}
We present the ab-initio thermoelastic properties of body-centered cubic molybdenum under extreme conditions obtained within the quasi-harmonic approximation including both the vibrational and the electronic thermal excitations contributions to the free
energy. The quasi-harmonic temperature dependent elastic constants are calculated and compared with existing experiments and with the quasi-static approximation. 
We find that the quasi-harmonic approximation allows a much better interpretation of the experimental data confirming the trend found previously in other metals.
Using the Voigt-Reuss-Hill average we predict the compressional and shear sound velocities of polycrystalline molybdenum as a function of pressure for several temperatures which might be accessible in experiments. 
\end{abstract}

\maketitle

\section{\label{sec:level1}Introduction}
Molybdenum, as a refractory $4d$ transition metal in the same group of tungsten, finds several applications, pure or in alloys with other metals, for its high melting point, mechanical properties, and corrosion resistance. Its thermodynamic properties have been studied by several authors, both experimentally and by theory.~\cite{ming_isothermal_1978,dewaele_compression_2008,zhao_thermoelastic_2000,litasov_thermal_2013_2,huang_thermal_2016,hixson_shock_1992,nix_thermal_1942,bodryakov_correlation_2014,zeng_lattice_2010,zeng_density_2011,zeng_dynamical_2014,wang_calculated_2000,cazorla_ab_2007}

Among the thermodynamic properties, the elastic constants (ECs) are key parameters that determine the mechanical stability, the velocity of sound waves, and the stress response to external strains. In this regard, particularly useful for applications is the pressure and temperature dependence of the ECs. In molybdenum, the temperature dependent bulk modulus and ECs have been measured by ultrasonic technique at room pressure.~\cite{featherston_elastic_1963,bolef_elastic_1962,dickinson_temperature_1967,bujard_elastic_1981} Data are
available almost until melting ($2898$ K).~\cite{bujard_elastic_1981} Pressure derivatives of the ECs are known at room temperature~\cite{katahara_pressure_1979}
and compressional and shear sound velocities in polycrystalline molybdenum have been measured up to $120$ kbar at room temperature.~\cite{liu_experimental_2009} A simultaneous measurement of the density allow to derive from these data the bulk and shear moduli at high pressure. However, information on pressure dependent elasticity at high temperatures is still missing in the literature.

Theoretically temperature dependent ECs of molybdenum have been calculated within the quasi-static approximation (QSA) at room pressure,~\cite{Wang_2010} while the pressure dependent ECs have been calculated 
by Ko\v ci et al.~\cite{koci_elasticity_2008} at
zero temperature.
The measurements on polycrystalline molybdenum have been modeled by ab-initio calculations~\cite{liu_experimental_2009} and the ECs along the Hugoniot together with the corresponding compressional and shear sound velocities have been calculated within the QSA.~\cite{zeng_density_2011}
Due to the time-consuming phonon calculations for deformed configurations of metallic systems needed for the ab-initio quasi-harmonic (QHA) ECs no paper has addressed these quantities for molybdenum so far.

In recent years, a workflow for the calculation of the QHA ECs within density functional theory (DFT) has been fully integrated in the \texttt{thermo\_pw} software,~\cite{dal_corso_thermo_pw_2022,malica_quasi-harmonic_2020} continuously improved, and applied to several solids (aluminum, silicon, BAs, copper, silver, gold, palladium, platinum, and tungsten).~\cite{malica_temperature-dependent_2019,malica_temperature_2020,malica_quasi-harmonic_2021,gong_pressure_2024} In many metals, it has been found that the QHA predicts the temperature dependence of the ECs accurately, much more than the QSA.~\cite{malica_quasi-harmonic_2020,pham_finite-temperature_2011}
The $0$ K values instead are similar and might differ from experiment, with 
differences that depend on the exchange and correlation functional and are
usually within $10$ \%.

In this paper we apply this technique to molybdenum.
We present a comparison of the temperature dependence of the QSA and the QHA ECs and use the calculated 
QHA adiabatic ECs to
predict the temperature and pressure dependence 
of the bulk and shear moduli of polycrystalline
molybdenum as well as its compressional and shear sound velocities, providing a theoretical prediction that could be useful in future investigations of the thermoelastic properties at high pressure and temperature.

\begin{table}[h]
  \caption{The equilibrium lattice constants ($a_0$), the bulk moduli ($B_T$) and the pressure derivatives of the bulk moduli ($B_T^{\prime}$) of molybdenum calculated in this work compared with previous calculations and with experiment.}\label{table:1}%
\begin{ruledtabular}
\begin{tabular}{cccccc}
 & & T  &$a_0$  & $B_T$  & $B_T^{\prime}$  \\
     & &  (K) & (a.u.) & (kbar) &   \\
\hline
This study &LDA & 0 & 5.884 & 2949 & 4.00  \\ 
  & & 295 & 5.894 & 2874 & 4.09  \\
 &PBEsol & 0 & 5.916 & 2826 & 4.02  \\
   & & 295 & 5.926 & 2753 & 4.10 \\
 &PBE & 0 & 5.975 & 2617 & 4.08 \\
   & &295  & 5.986 & 2543 & 4.16  \\
 \newline  \\
 Calc.~\cite{zeng_lattice_2010}& PW91 & 0& 5.988 & 2666 & 4.42 \\
 Calc.~\cite{zeng_density_2011}& PBE & 0& 5.996 & 2633 & 4.21 \\
 Calc.~\cite{koci_elasticity_2008}& PBE & 0& 6.001 & 2610 & 4.5 \\
 Calc.~\cite{wang_calculated_2000}\footnotemark[1] & PBE & 0& 5.992 &  &  \\
 Calc.~\cite{haas_calculation_2009} & LDA & 0& 5.888 &  &  \\
 & PBEsol & 0& 5.920 &  &  \\
 & PBE & 0& 5.979 &  &  \\
  Calc.~\cite{dewaele_compression_2008} & LDA & 0& 5.880 & 3010 & 3.99 \\
  & & 298\footnotemark[2] & 5.891 & 2950 & 4.01 \\
 & PBEsol & 0& 5.914 & 2870 & 4.02 \\ 
 &  & 298\footnotemark[2] & 5.925 & 2800 & 4.05 \\
 & PBE & 0& 5.981 & 2620 & 4.14 \\
 & & 298\footnotemark[2] & 5.993 & 2550 & 4.17 \\
  \newline  \\
 Model~\cite{litasov_thermal_2013_2}&  &300& 5.945 & 2600 & 4.21 \\ Model~\cite{sokolova_self_2013}&  &300& 5.944 & 2605 & 4.05 \\
 \newline  \\
 Expt.~\cite{dewaele_compression_2008}\footnotemark[1] & & 300 & 5.944 & 2610 & 4.06 \\
 Expt.~\cite{ming_isothermal_1978}& &300 & 5.951 & 2608 & 4.46 \\ 
 Expt.~\cite{featherston_elastic_1963}& & 0 &  & 2653 &  \\ 
 Expt.~\cite{katahara_bcc_1976}& & & & 2610 & 4.65\footnotemark[3]/3.95\footnotemark[4] \\ 
\end{tabular}
\end{ruledtabular}
\footnotetext[1]{These data are used to calculate the equations of state which we report in Fig.~S2 and Fig.~S3 of the supplementary material.}
\footnotetext[2]{Values estimated using a Debye model.}
\footnotetext[3]{Ultrasonic experiment.}
\footnotetext[4]{Shock Wave experiment.}

\end{table}

\section{Theory and computational parameters}\label{sec3}
Within the QHA, the Helmholtz free energy of a cubic solid is a function
of temperature $T$ and (unit cell) volume $V$. It can be written as the sum
of three contributions:
\begin{equation}
F(V,T)= U(V) + F_{ph}(V,T) + F_{el}(V,T),
\label{eq:free_ener}
\end{equation}
where $U(V)$ is the static energy, computed by DFT, $F_{el}(V,T)$ is the electronic thermal excitations contribution to the free energy, and $F_{ph}(V,T)$ is the vibrational free energy written in terms of the phonon frequencies
${\omega}_{\eta}({\bf q},V)$:

\begin{eqnarray} \label{equ4}  
F_{vib}(V, T) & =& \frac{1}{2N} \sum_{\mathbf q \eta} \hbar \omega_{\eta} \left(\mathbf q,  
V \right) \nonumber\\  
& +& {\frac{1}{N \beta}} \sum_{\mathbf q \eta} \ln \left[1 - \exp \left(- \beta \hbar \omega_{\eta}(\mathbf q, V)\right) \right].  
\end{eqnarray} 
$N$ is the total number of cells in the solid
(equal to the number of {\bf q} vectors in 
the sum) and $\hbar$ is the reduced Planck's constant. $\mathbf q$ are the phonon wave vectors and the subscript $\eta$ indicates the different vibrational modes.

The contribution of the electronic thermal excitations to the free energy is calculated within the rigid bands approximation~\cite{malica_quasi-harmonic_2021}
as $F_{el}=U_{el}-TS_{el}$. 
$U_{el}$ is electronic excitation contribution to the energy given by
\begin{equation}
U_{el}=\int_{-\infty}^\infty E N(E) f(E,T,\mu) dE - 
\int_{-\infty}^{E_F} E N(E) dE,
\end{equation}
where $E_F$ is the Fermi energy, $N(E)$ is
the electronic density of states, 
$f(E,T,\mu)$ are the Fermi-Dirac occupations,
$\mu$ is the chemical potential, and
$S_{el}$ is the electronic entropy given by:
\begin{eqnarray}
S_{el}=&-&k_B\int_{-\infty}^\infty \Bigg[ f(E,T,\mu) \ln f(E,T,\mu) 
 \nonumber \\ &+& (1 - f(E,T,\mu)) \ln (1 - f(E,T,\mu)) \Bigg] N(E) dE. 
\end{eqnarray}
For a certain number of geometries $N_V$, a fourth order Birch-Murnaghan equation is applied to interpolate the $U(V)$ at cell volumes $V_i$, $i=1,2, ..., N_V $. The vibrational and electronic free energies are fitted at each temperature by a fourth degree polynomial as a functions of V.
The computational details for the volume thermal expansion $\beta_V$, the isobaric heat capacity $C_p$, the isoentropic bulk modulus $B_S$ and the average Gr\"uneisen parameter are given in the supplementary material.

The isothermal ECs are calculated from the second strain derivatives of the free energy:  
\begin{equation}  
\tilde C^T_{ijkl}= {1\over V} { \partial^2 F \over \partial \varepsilon_{ij}  
\partial \varepsilon_{kl}} \Bigg |_T,  
\label{tdec}  
\end{equation}  
correcting for finite pressure effects to obtain the stress-strain ECs:~\cite{barron_second-order_1965}  
\begin{equation}  
C^T_{ijkl} = \tilde C^T_{ijkl} + {p \over 2} \left(2 \delta_{i,j} \delta_{k,l}  
- \delta_{i,l} \delta_{j,k} - \delta_{i,k} \delta_{j,l}  \right).  
\end{equation}  
The second derivatives of the free energy are calculated as described in Ref.~\cite{malica_quasi-harmonic_2020} taking a subset of the volumes $V_i$ as equilibrium configurations. The ECs at any other volume at temperature $T$ and pressure $p$ are obtained by interpolation by a fourth-degree polynomial. Adiabatic ECs are calculated from the isothermal ones as:  
\begin{equation}  
C^S_{ijkl}=C^T_{ijkl} + {T V b_{ij} b_{kl} \over C_V},  
\end{equation}  
where $b_{ij}$ are the thermal stresses:  
\begin{equation}  
b_{ij} = - \sum_{kl} C^T_{ijkl} \alpha_{kl},  
\end{equation}  
and $C_V$ is the isochoric heat capacity (See the supplementary material). For a cubic system, the linear thermal expansion tensor is diagonal and from the volume thermal expansion $\beta_V$ we get: $\alpha_{kl} = \delta_{kl} \beta_V/3 $. With the knowledge of the adiabatic ECs, using the Voigt-Reuss-Hill approximation, we compute the polycrystalline average of the bulk modulus $B_S$, of the shear modulus $G_S$, of the Young's modulus $E_S$, and of the Poisson's ratio $\nu_S$ (see the \texttt{thermo\_pw} manual for the expressions used). Finally, the compressional and the shear sound velocities are given by:  
\begin{equation}  
V_P= \sqrt{ B_S + {4 / 3} G_S \over \rho },  
\label{sound_p}  
\end{equation}  
\begin{equation}  
V_G= \sqrt{ G_S \over \rho},  
\label{sound_g}  
\end{equation}  
where $\rho$ is the density. Note that these equations also hold with pressure as long as $B_S$ and $G_S$ are computed from the stress-strain ECs.~\cite{wallace}

The temperature and pressure dependent thermodynamic properties and ECs are calculated by the open source software \texttt{thermo\_pw}, which has been  discussed in previous publications.~\cite{dal_corso_elastic_2016,pal2,malica_temperature-dependent_2019,malica_quasi-harmonic_2020,malica_temperature_2020,malica_quasi-harmonic_2021,gong_pressure_2024}

The calculations presented in this work are done by using DFT as implemented in the Quantum ESPRESSO (QE) package.~\cite{qe1, qe2} The exchange and correlation functionals are the LDA~\cite{lda} and the generalized gradient approximations PBEsol~\cite{pbesol} and PBE.~\cite{pbe}

We employ the projector augmented wave (PAW) method~\cite{paw} and a plane-wave basis with pseudopotentials from \texttt{pslibrary}.~\cite{corso_dalcorsopslibrary_2022}
We use \texttt{Mo.pz-spn-kjpaw\_psl.1.0.0.UPF}, \texttt{Mo.pbesol-spn-kjpaw\_psl.1.0.0.UPF}, and \texttt{Mo.pbe-spn-kjpaw\_psl.1.0.0.UPF} for LDA, PBEsol, and PBE, respectively. These pseudopotentials have the $4s$, $4p$, $4d$, and $5s$ states in the valence, while the other states are frozen in the core and accounted for by the nonlinear core correction.~\cite{louie_nonlinear_1982} 
The lattice constants of 14 reference geometries from $4.784$ a.u. to $6.084$ a.u. with LDA, from
$4.8162$ a.u. to $6.1162$ a.u. for PBEsol, and from $4.8922$ a.u. to $6.1922$ a.u. with PBE with an interval of $0.1$ a.u. between geometries have been chosen to calculate the free energies. For the wave functions cutoffs, we use $100$ Ry, $90$ Ry, $120$ Ry while for the charge density we use $400$ Ry, $360$ Ry, $480$ Ry, for LDA, PBEsol, and PBE, respectively. The Fermi surface has been dealt with by the smearing approach of Methfessel and Paxton~\cite{mp} with a smearing parameter $\sigma = 0.02$ Ry. With this smearing, the Brillouin zone integrals converge with a $40 \times 40 \times 40$ {\bf k}-point mesh.

For $6$ reference geometries (with $i=1$, $4$, $7$, $11$, $12$, $13$), temperature dependent ECs are calculated by $3$ strain types that lead to a body-center cubic, a centered tetragonal, and a rhombohedral strained lattices. Each strain type is sampled by 6 strains, from $\epsilon=-0.15$ to $\epsilon=0.15$ with stepsize $\delta_{\epsilon}= 0.05$. A thicker {\bf k}-point mesh of $45 \times 45 \times 45$ is employed on strained configurations. Each one of the $108$ strained configurations requires calculations of phonon frequencies and electronic density of states. Phonon frequencies
are calculated by density functional perturbation theory (DFPT)~\cite{rmp, dfptPAW} getting the dynamical matrices on a $8 \times 8 \times 8$ \textbf{q}-point grid. These dynamical matrices have been Fourier interpolated on a $200 \times 200 \times 200$ \textbf{q}-point mesh to evaluate the free-energy and the thermodynamic quantities.
The calculations are all performed on the Leonardo supercomputer at CINECA with a GPU version of \texttt{thermo\_pw} that optimizes some routines of QE for problems with dense {\bf k}-points sampling in metallic systems.~\cite{gong_dalcorso_opt}

\begin{figure}
\centering
\includegraphics[width=\linewidth]{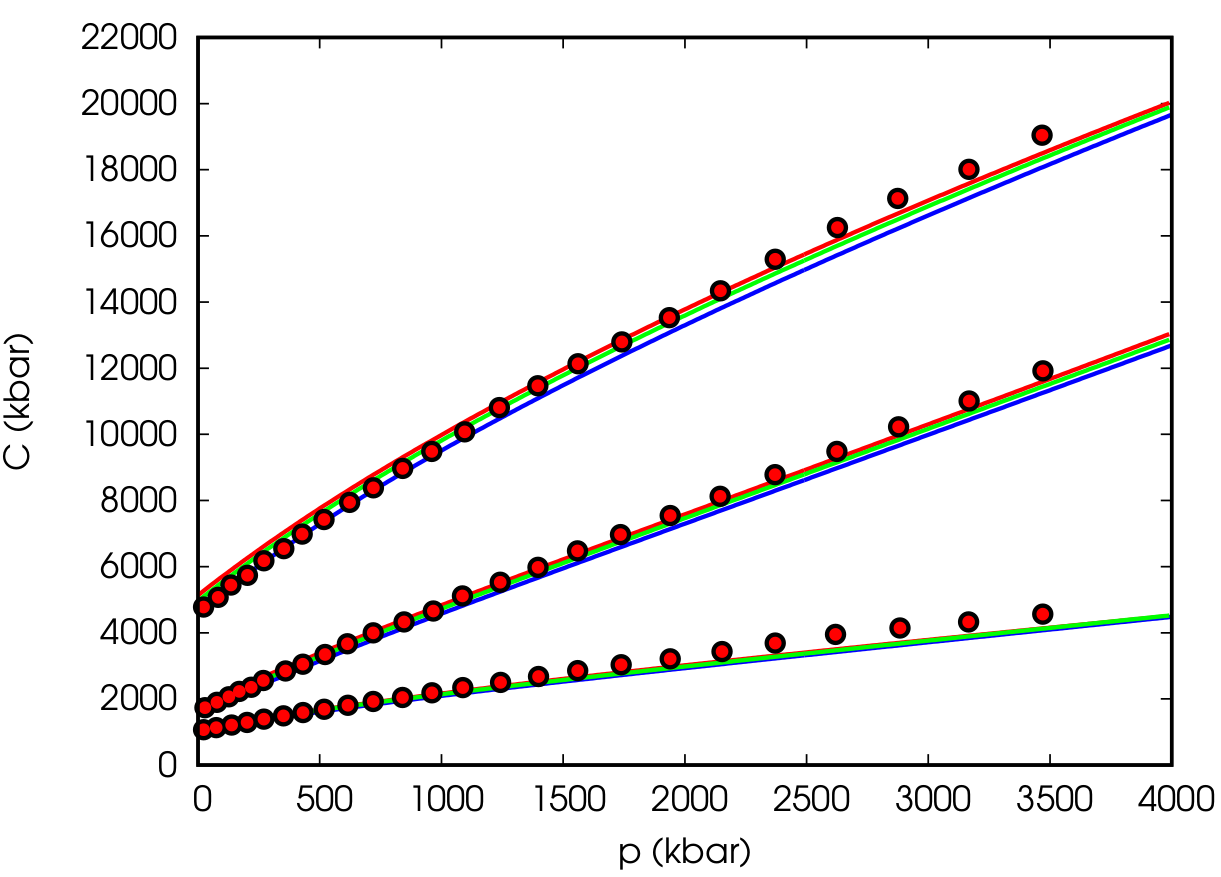}
\caption{Elastic constants as a function of pressure calculated within LDA (red lines),  PBEsol (green lines) and PBE (blue lines) compared
with the PBE results of Ref.~\cite{koci_elasticity_2008}.} 
\label{fig:elastic_p}
\end{figure}

\begin{figure}
\centering
\includegraphics[width=\linewidth]{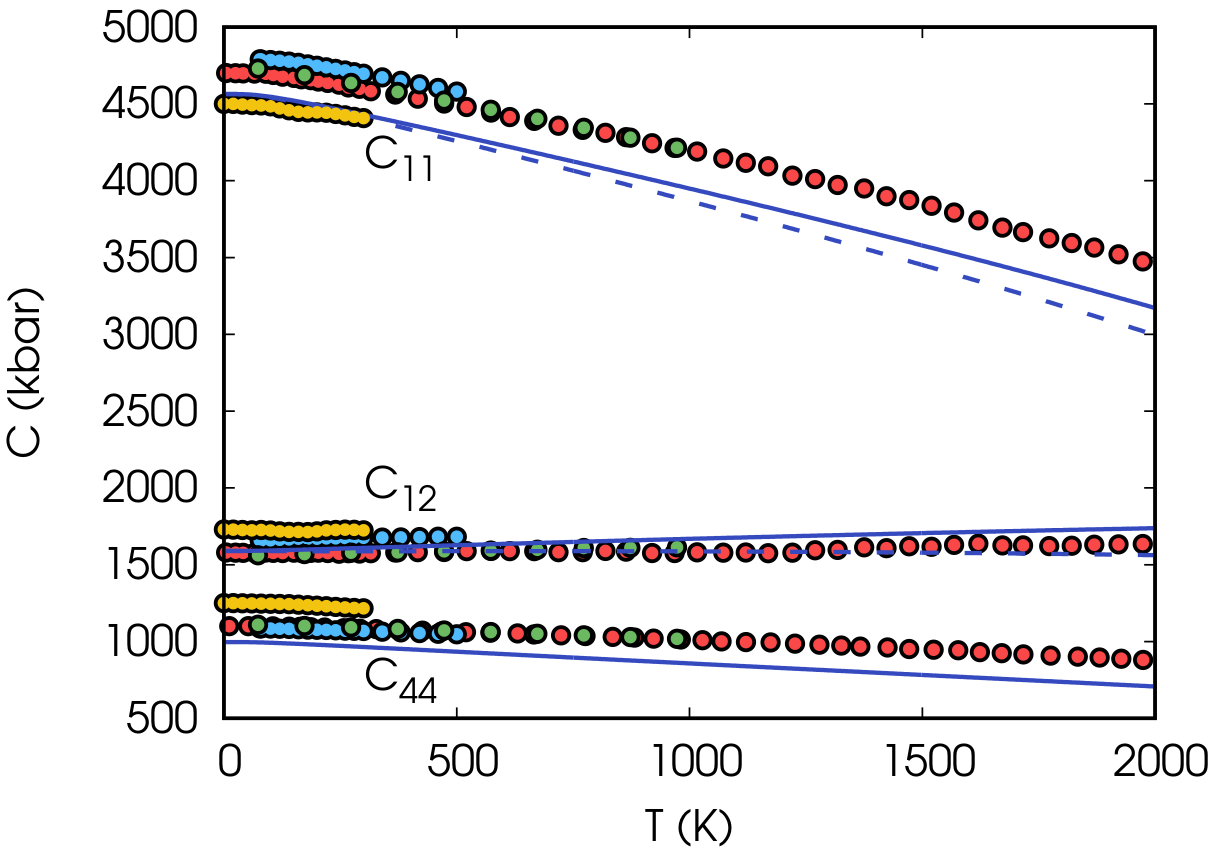}
\caption{Quasi-harmonic isothermal (dashed  lines) and adiabatic (solid line) elastic constants $C_{11}$, $C_{12}$ and $C_{44}$ as a function of temperature compared with experimental adiabatic data from Ref.~\cite{featherston_elastic_1963}
(yellow circles), Ref.~\cite{dickinson_temperature_1967} (green circles), Ref.\cite{bolef_elastic_1962} (blue circles), and Ref.\cite{bujard_elastic_1981} (red circles).} 
\label{fig:el_cons_t}
\end{figure}

\begin{figure}
\centering
\includegraphics[width=\linewidth]{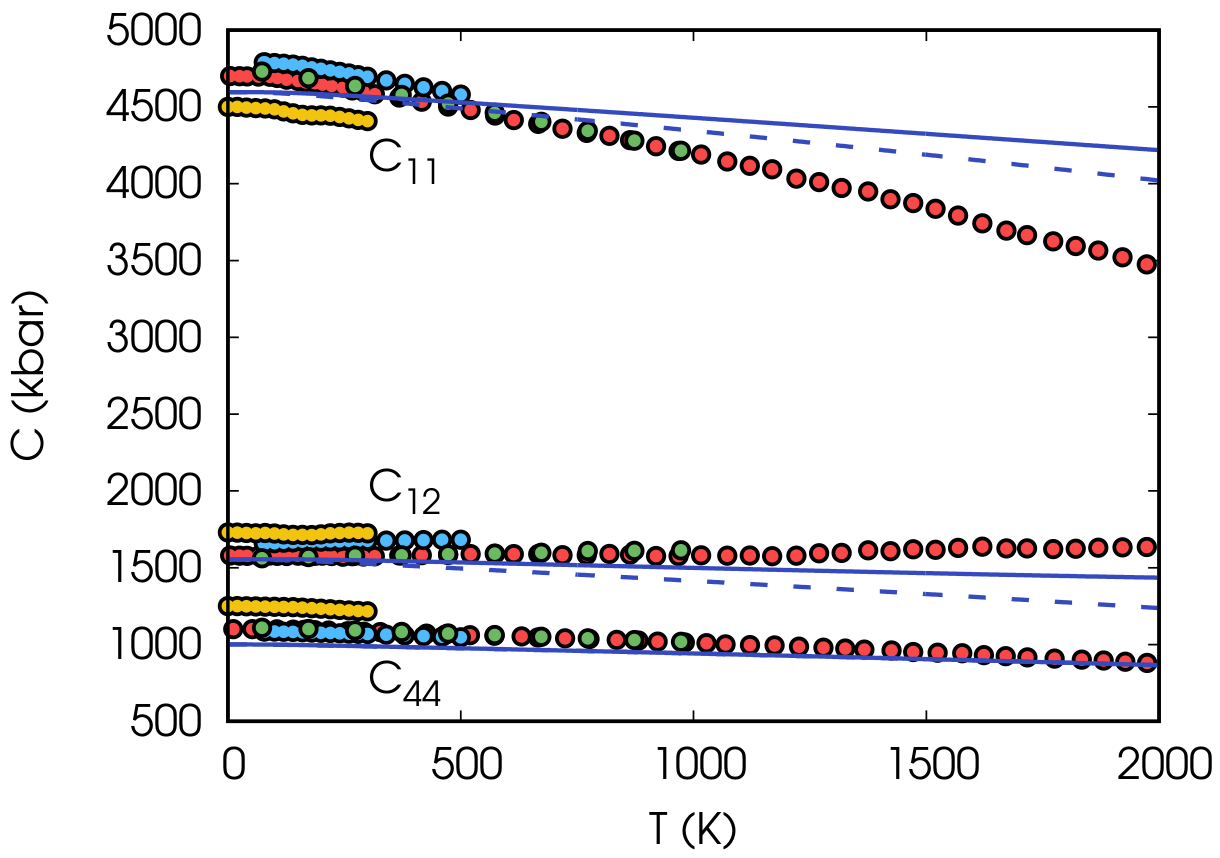}
\caption{Quasi-static isothermal (dashed lines) and adiabatic (solid line) elastic constants $C_{11}$, $C_{12}$, and $C_{44}$ as a function of temperature compared with adiabatic experimental data from Ref.~\cite{featherston_elastic_1963}
(gold circles), Ref.~\cite{dickinson_temperature_1967} (green circles), Ref.\cite{bolef_elastic_1962} (blue circles), and Ref.\cite{bujard_elastic_1981} (red circles).} 
\label{fig:elastic_t_qsa}
\end{figure}

\section{Results and Discussion}\label{sec5}
In Table~\ref{table:1}, the equilibrium lattice constants, bulk moduli, and pressure derivatives of the bulk moduli of molybdenum obtained as parameters of a fourth-order Birch-Murnaghan interpolation of the static energy $U(V)$ are listed together with a few selected values from previous calculations and experiment.
Our PAW LDA, PBEsol, and PBE values of the lattice constant differ by less than $0.1 \%$ from the all-electron values reported in Ref.~\cite{haas_calculation_2009}. 
With respect to experiment ($a=5.936$ a.u.
at $0$ K) the LDA, PBEsol, and PBE errors are $-0.9 \%$, $-0.3 \%$ and $0.7 \%$, with LDA and PBEsol below experiment and
PBE above. For the bulk modulus these
errors become $10 \%$ (LDA),  $7 \%$ (PBEsol), and $-1 \%$ (PBE) with respect to the 0 K value $2653$ kbar.~\cite{featherston_elastic_1963}

\begin{table*} 
  \caption{\label{table:2} The $0$ K elastic constants calculated with the different functionals compared with experiment and one previous calculation. $B$, $E$, $G$, and $\nu$ are the bulk modulus, the Young's modulus, the shear modulus, and the Poisson's ratio, respectively.}
\begin{ruledtabular}
\begin{tabular}{cccccccccc}
 & T &$a_0$  & $C_{11}$  & $C_{12}$ & $C_{44}$ & $B$ & $E$ & $G$ & ${\nu}$ \\
  & (K) & (a.u.) & (kbar)  & (kbar) & (kbar) & (kbar) & (kbar) & (kbar) &  \\
\hline 
 LDA & 0 & 5.884  & 5183  &   1815 &  1094  & 2938 & 3402 & 1301  & 0.307 \\ 
 PBEsol & 0 & 5.916 & 4976  & 1727 &  1081 & 2810  & 3318 & 1273  & 0.303 \\
 PBE & 0 &5.974 & 4637 & 1589 & 1016 &   2605 & 3111  & 1196  &  0.301 \\
 PW91~\cite{zeng_lattice_2010} & 0 &5.988 & 4723 & 1604 & 1060 & 2644 & 3211  & 1237  & 0.297  \\
Expt.~\cite{featherston_elastic_1963} & 0 & & 4500.2 &1729.2 & 1250.3 &  2653  & 3358  & 1303  & 0.289  \\
Expt.~\cite{dickinson_temperature_1967} & 273.15 & & 4637 &1578 & 1092 & 2598  & 3232  & 1250  & 0.293  \\
Expt.~\cite{dickinson_temperature_1967} (Extrapolated) & 0 & & 4800&1558 &1124  & 2639  & 3354  & 1302  & 0.288  \\
Expt.~\cite{bolef_elastic_1962} & 300 & & 4696 & 1676 & 1068 & 2683  & 3194  & 1227  & 0.302  \\
Expt.~\cite{bolef_elastic_1962} (Extrapolated) & 0 & & 4832 & 1656 & 1100 & 2715  & 3306  & 1275 & 0.297  \\
Expt.~\cite{katahara_pressure_1979} & 300 & &4648 & 1616 & 1089 & 2627  & 3222  & 1244 & 0.296  \\
Expt.~\cite{katahara_pressure_1979} (Extrapolated) & 0 & & 4784 & 1596 & 1121    & 2659  & 3334 & 1291 & 0.291 \\
Expt.~\cite{liu_experimental_2009} & 300 & & &  &  & 2607  &   & 1251 &   \\
 \end{tabular}
\end{ruledtabular}
 
\end{table*} 

In Table~\ref{table:2}, we report the values of the ECs $C_{11}$, $C_{12}$, and $C_{44}$ calculated with the three functionals together with the values of the bulk modulus, Young's modulus, shear modulus, and Poisson's ratio of polycrystalline molybdenum calculated using the Voigt-Reuss-Hill approximation. The temperature dependent ECs have been measured in Refs.~\cite{bolef_elastic_1962,featherston_elastic_1963,dickinson_temperature_1967,bujard_elastic_1981}. Although there
is not perfect agreement among these data, the $0$ K values of
Ref.\cite{dickinson_temperature_1967,bujard_elastic_1981,katahara_pressure_1979} are quite close to each other. Taking as a reference the values of Ref.~\cite{katahara_pressure_1979} extrapolated to $0$ K, by adding the theoretical difference between $0$ K and $300$ K, we
find that the LDA errors for $C_{11}$, $C_{12}$, and $C_{44}$ are $399$ kbar ($8$ \%), $219$ kbar ($14$ \%) 
and $-27$ kbar ($-2$ \%) while the PBE errors are 
$-147$ kbar ($-3$ \%), $-7$ kbar ($-0.4$ \%) and $-105$ kbar
($-9$ \%). PBEsol has
errors $192$ kbar ($4$ \%), $131$ kbar ($8$ \%), and $-40$ kbar ($-4$ \%) smaller than LDA, but bigger than PBE. 
The PBE values of $C_{11}$ and $C_{12}$, and hence of its bulk modulus, are
the closest to experiment. Since the calculation of the TDEC is computationally heavy, we calculated them using only this functional. Actually, as
shown in Ref.~\cite{malica_quasi-harmonic_2021} for several metals 
and as we confirmed in a recent study of 
tungsten,~\cite{gong_ab_2024} different functionals give different $0$ K values of
the ECs but the temperature and
pressure dependence is almost 
independent from the functional.

\begin{figure}
\centering
\includegraphics[width=\linewidth]{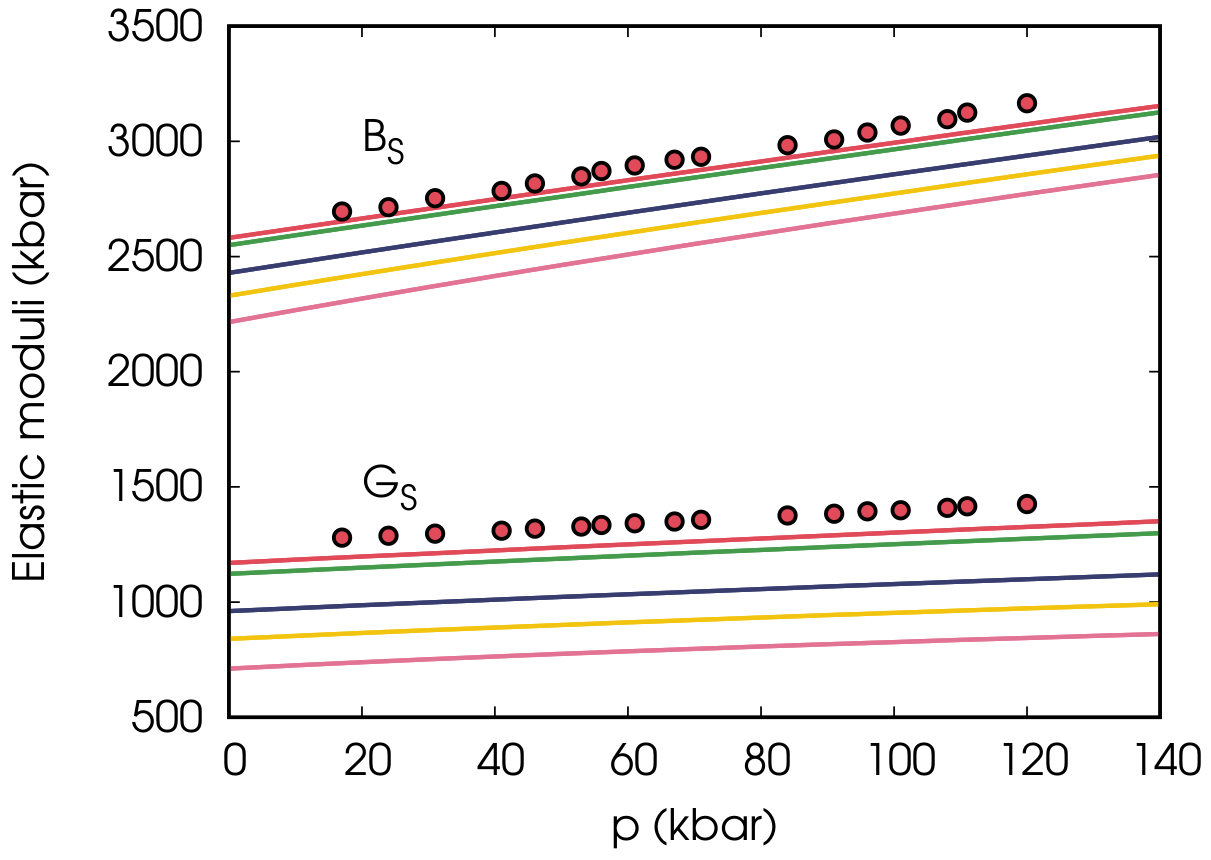}
\caption{Adiabatic bulk and shear modulus of polycrystalline molybdenum against pressure computed at $5$ K (red line), $300$ K
(green line), $1000$ K (blue line), $1500$ K (yellow line), and $2000$ K (pink line), compared with
room temperature experimental values of Ref.~\cite{liu_experimental_2009} (red circles).} 
\label{fig:elastic_moduli_mo}
\end{figure}

\begin{figure}
\centering
\includegraphics[width=\linewidth]{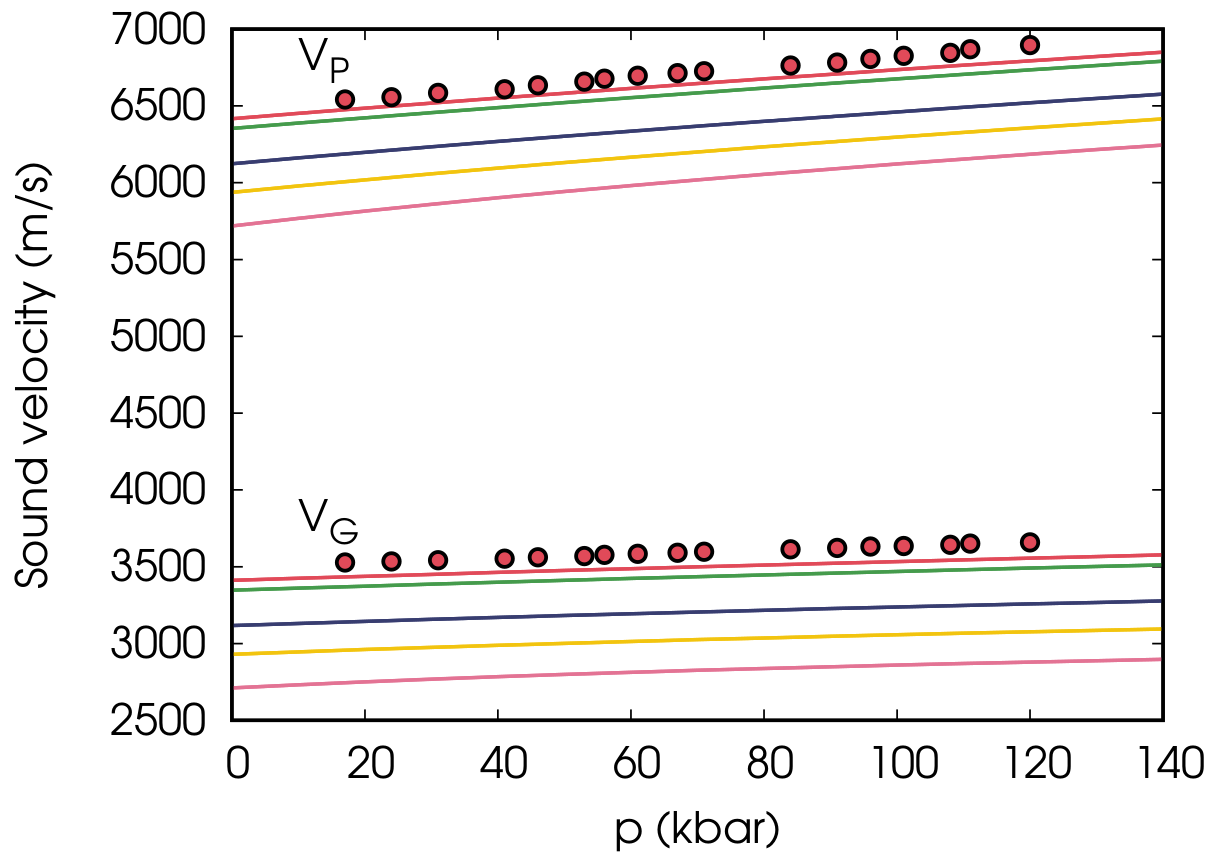}
\caption{Compressional and shear sound velocities of polycrystalline molybdenum against pressure at $5$ K (red line), $300$ K
(green line), $1000$ K (blue line), $1500$ K (yellow line), and $2000$ K (pink line), compared with
room temperature experimental values of Ref.~\cite{liu_experimental_2009} (red circles).} 
\label{fig:sound_velocity_mo}
\end{figure}

In Fig.~\ref{fig:elastic_p} we present the pressure 
dependent ECs at $0$ K calculated with the three functionals and compare them with the PBE results of Ko\v ci et al..~\cite{koci_elasticity_2008} There is a reasonable agreement between the two calculations expecially at low pressure. At $3000$ kbar our ECs are smaller than those of Ko\v ci et al. but quite close to them.
At zero pressure, the pressure derivatives of the ECs are: 
${dC_{11} \over dp} =5.8$,
${dC_{12} \over dp} =3.3$,
${dC_{44} \over dp} =1.3$, for 
all three functionals to be compared to the experimental values:~\cite{katahara_pressure_1979} 
${dC_{11} \over dp} =6.41$,
${dC_{12} \over dp} =3.45$, and
${dC_{44} \over dp} =1.396$. 

We show in Fig.~\ref{fig:el_cons_t} the QHA isothermal and
adiabatic ECs compared with the adiabatic
experimental values. Keeping into account the zero point motion on both the lattice constant and on the ECs themselves, we find $C_{11}=4564$ kbar, $C_{12}=1589$ kbar, and $C_{44}=996$ kbar at $4$ K, while computing the ECs from the strain derivatives of the energies at the lattice constant expanded by zero point motion effects within the QSA we get $C_{11}=4593$ kbar, $C_{12}=1565$ kbar, and $C_{44}=996$ kbar.

As can be seen from Fig.~\ref{fig:el_cons_t}, there is a good
agreement between our calculated temperature dependence and the experimental data. From $24$ K and $2022$ K, the experimental values~\cite{bujard_elastic_1981} decrease of $1264$ kbar ($27$ \%), $-63$ kbar ($-4$ \%), and $231$ kbar ($21$ \%) for $C_{11}$, $C_{12}$, and $C_{44}$ respectively, while our values decrease by $1400$ kbar ($31$ \%), $-141$ kbar ($-9$ \%), and $287$ kbar ($29$ \%).
In particular the increase of $C_{12}$ with temperature is found also in our QHA calculation, slightly
overestimated with respect to experiment.

For comparison we show in Fig.~\ref{fig:elastic_t_qsa}
the ECs calculated with the PBE functional within the QSA which are in good agreement with those calculated in Ref.~\cite{Wang_2010}.
In this case, from $4$ K to $2000$ K, the decreases of $C_{11}$, $C_{12}$, and
$C_{44}$ are $380$ kbar ($8$ \%), $127$ kbar ($8$ \%), and 
$132$ kbar ($13$ \%). The decrease of $C_{11}$, and $C_{44}$ is much smaller than in experiments (and within QHA)
while $C_{12}$ decreases with temperature instead of increasing as in experiment.
We can understand this behaviour using
the QHA ECs calculated at fixed volume 
that do not contain any thermal expansion effect. For these ECs $C_{11}$ and $C_{44}$
decrease with temperature, while $C_{12}$
increases. Since QSA has only the effect of thermal espansion for $C_{11}$ and $C_{44}$ it misses the QHA contribution
that give a larger decrease, while for $C_{12}$ it has no increasing QHA term.
The QHA predicts an almost
constant $C_{12}$ that is the results
of the cancellation between the decrease
due to thermal expansion and the increase
due to the use of the free energy derivatives instead of the energy derivatives.

Using the QHA ECs we have calculated the properties of polycrystalline molybdenum.
In Fig.3 we show the pressure dependence of the bulk modulus and of the shear modulus in the range of pressures (up to 140 kbar) measured in
Ref.~\cite{liu_experimental_2009}. In addition to
the $300$ K calculation (green line), which can be compared with experiment, we show our predictions for $4$ K, $1000$ K, $1500$ K and $2000$ K.
We can see that the derivatives of the bulk and shear modulus with respect to pressure are well followed by our curves. Our values at $300$ K are ${dB_S \over dp} =4.24$ and
${dG_S \over dp} =1.33$ against experimental values (obtained by a linear fit) ${dB_S \over dp} = 4.54$  and ${dG_S \over dp} =1.5$, respectively. These data are in agreement with the
experimental values reported in Ref.~\cite{katahara_pressure_1979}:
${dB_S \over dp} = 4.44$  and ${dG_S \over dp} =1.43$
and with the $0$ K PBE theoretical results of Ref.~\cite{liu_experimental_2009} ${dB \over dp} = 4.4$  and ${dG \over dp} =1.7$.
Regarding the temperature dependence of 
the adiabatic bulk and shear modulus we
find the following derivatives at $298$ K:
${dB_S \over dT} = -0.15$ kbar/K  and ${dG_S \over dT} = -0.22$ kbar/K.

Finally using Eq.~\ref{sound_p} and Eq.~\ref{sound_g}, we
computed the compressional and shear sound velocities
as a function of pressure for the same set of temperatures used in the previous picture. They are
presented in Fig.~\ref{fig:sound_velocity_mo}.
Even in this case the pressure dependence of the
sound velocity at $300$ K is well reproduced by the calculation and the other curves are our prediction. 

\section{Conclusions}
We presented the temperature and pressure dependent thermoelastic properties of molybdenum calculated by the \texttt{thermo\_pw} software
and PAW pseudopotentials. 
We find that the QHA predicts the temperature dependence of the ECs in much better agreement with experiments than the QSA.
Furthermore we have used the QHA ECs
to compute the pressure dependent compressional and shear sound velocities in polycrystalline 
molybdenum (and the corresponding bulk and shear moduli). In addition to the calculation at $300$ K that is in good agreement with the experimental results of Liu et al.,~\cite{liu_experimental_2009} we have calculated the low temperature ($4$ K) and the high temperature
($1000$ K, $1500$ K, and $2000$ K) pressure dependent curves, hoping that these calculations will stimulate an experimental investigation of these quantities.

For the sake of completeness, the phonon dispersions, the p-V equation of state at $300$ K and $2000$ K,
the temperature dependent volume thermal expansion, 
the isobaric heat capacity, the adiabatic bulk modulus, and the average Gr\"uneisen parameter for $0$ kbar, $1000$ kbar, $2000$ kbar, and  $3000$ kbar have been calculated by the LDA, PBEsol, and PBE functionals, but since they are already available in the literature, we have 
moved them to the supplementary material.

\section{Supplementary material}
Supplementary material that presents the calculated
thermodynamic properties and a few tests on some
numerical parameters are available at the following
link:

\begin{acknowledgments}
This work has been supported by the Italian MUR (Ministry of University and Research) through the National Centre for HPC, Big Data, and Quantum Computing (grant No. CN00000013). Computational facilities have been provided by SISSA through its Linux Cluster, ITCS, and the SISSA-CINECA 2021-2024 
Agreement. Partial support has been received from the European Union through the MAX ``MAterials design at the eXascale'' Centre of Excellence for Supercomputing applications (Grant agreement No. 101093374, co-funded by the European High Performance Computing joint Undertaking (JU) and participating countries 824143). X. Gong acknowledges the support received in the framework of the Joint Research Agreement for Magnetic Confinement Fusion between Eni and CNR.

\end{acknowledgments}

\bibliographystyle{apsrev4-1}
\bibliography{aipsamp}


\end{document}



\title{Supplementary material for: Ab initio quasi-harmonic thermoelasticity of molybdenum under high temperature and pressure} 



\author{Xuejun Gong}
\email[]{xgong@sissa.it}
\affiliation{International School for Advanced Studies (SISSA), Via Bonomea 265, 34136, Trieste, Italy}
\affiliation{IOM - CNR, Via Bonomea 265, 34136, Trieste, Italy}
\author{Andrea Dal Corso}
\email[]{dalcorso@sissa.it}
\affiliation{International School for Advanced Studies (SISSA), Via Bonomea 265, 34136, Trieste, Italy}
\affiliation{IOM - CNR, Via Bonomea 265, 34136, Trieste, Italy}


\date{\today}

\begin{abstract}
This supplementary material documents the phonon dispersions, thermal expansion, isobaric heat capacity, adiabatic bulk modulus, and average Gr\"uneisen parameter of molybdenum under high temperature and pressure, calculated by \texttt{thermo\_pw} with three exchange
and correlation functionals (LDA~\cite{lda}, PBEsol~\cite{pbesol}, and PBE~\cite{pbe}). Available results from experiments and previous calculations are also marked on the figures for reference.
\end{abstract}

\pacs{}

\maketitle 



\section{Theory }
We start by summarizing the thermodynamic relationships used in this suplementary
material.

Considering $p$ as a fixed parameter, the minimization of the functional $G_p(V,T)=F(V,T)+pV$ with respect to the volume gives the equation of state (EOS): 
\begin{equation} 
p=-{\frac{\partial F(V,T)}{\partial V}}.
\label{eq:p} 
\end{equation} 
Hence the volume that minimizes $G_p(V,T)$ is the volume at pressure $p$ and temperature $T$: $V(p,T)$. 
Using $V(p,T)$ we obtain the volume thermal expansion $\beta_V(p,T)$ at pressure $p$ as: 
\begin{equation} 
\beta_V(p,T)= {\frac{1}{V(p,T)}} {\frac{\partial V(p,T)}{\partial T}} \Bigg |_p. 
\label{eq:beta} 
\end{equation} 
Using $p(V,T)$ (Eq.~\ref{eq:p}), we compute the isothermal bulk modulus as: 
\begin{equation} 
B_T(V,T) = - V { \frac{\partial p (V,T)} {\partial V}} \Bigg |_T = V { \frac{\partial^2 F (V,T)} {\partial V^2}} \Bigg |_T. 
\label{eq:bm} 
\end{equation} 

The part of the bulk modulus $B_T$ that derives from $U(V)$ is a parameter of the Birch-Murnaghan equation, while the thermal part which derives from $F_{ph} + F_{el}$ is calculated by a second volume derivative of the polynomial interpolation. 

The vibrational contribution to the isochoric heat capacity $C_{V,ph}(V,T)$ is calculated for the $N_V$ geometries by:  
\begin{equation} 
C_{V,ph}(V,T) = {\frac{1}{N}} \sum_{{\bf q},\eta} \hbar \omega_\eta({\bf q}, V)  
{\frac{\partial}{ \partial T}} \left[ \frac{1} { e^{\beta \hbar  
\omega_\eta({\bf q},V)} - 1} \right]. 
\label{eq:cv} 
\end{equation} 
The electronic contribution $C_{V,el}(V_i,T)$ is computed from the electronic density of states and is added to $C_{V,ph}(V_i,T)$ to obtain $C_V(V_i,T)=C_{V,ph}(V_i,T)+C_{V,el}(V_i,T)$. 
At each temperature, these quantities are fitted by a fourth-degree polynomial in $V$. 
The isobaric heat capacity is given by: 
\begin{equation} 
C_p(p,T) = C_V(V,T) + \beta_V^2(p,T) B_T(V,T) V T, 
\label{eq:cp} 
\end{equation} 
the adiabatic bulk modulus is: 
\begin{equation} 
B_S(p,T) = B_T(V,T) + {\frac{\beta_V^2(p,T) B^2_T(V,T) V T}{C_V(V,T)}}, 
\label{eq:bs} 
\end{equation} 
and the thermodynamic average Gr\"uneisen parameter is defined as: 
\begin{equation} 
\gamma(p,T) = {\frac{\beta_V(p,T) B_T(V,T) V}{C_V(V,T)}}. 
\label{eq:gamma} 
\end{equation} 

In the present supplementary material, in addition to $\beta_V(p,T)$, we show $C_p(p,T)$, $B_S(p,T)$, and $\gamma(p,T)$ for several pressures $p$, so when $V$ appears in Eqs.~\ref{eq:cp}-\ref{eq:gamma} we use $V(p,T)$.

\section{Results}

\subsection{Phonon dispersion}

\begin{figure}[H]
\centering
\includegraphics[width=\linewidth]{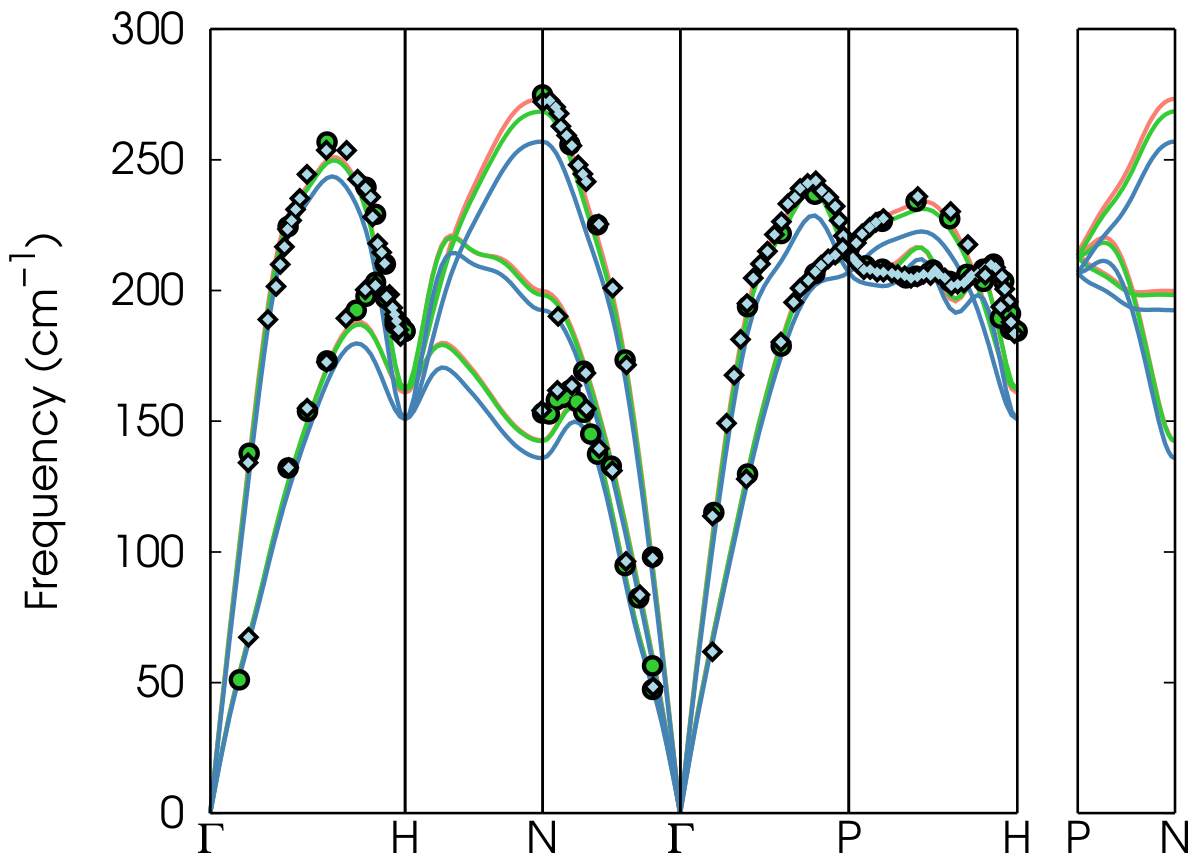}
\caption{Phonon dispersions interpolated at the $295$ K lattice constant. The LDA (red curves), PBEsol (green curve), and PBE (blue curve) results are compared with the experimental inelastic neutron scattering data measured at $295$ K (Ref.~\cite{powell_lattice_1968} blue diamonds and Ref.\cite{zarestky_temperature_1983} green circles).} 
\label{fig:ph_disp}
\end{figure}

In Fig.~\ref{fig:ph_disp}, we compare the theoretical phonon dispersions, interpolated at the $295$ K lattice constant obtained
accounting for zero point and thermal expansion effects (see the values in Table~1 of the paper), with inelastic neutron scattering data.~\cite{powell_lattice_1968,zarestky_temperature_1983} 
LDA and PBEsol give similar frequencies in good agreement with experiment except at the $H$ point where we find a relatively large error. PBE gives frequencies lower than LDA (as found in other solids~\cite{dalcorso_lessigreaterab_2013,gong_ab_2024}), and hence more distant from experiment. 
As for gold, platinum and tungsten, for molybdenum phonon dispersions, LDA or PBEsol are preferable to PBE. This conclusion
is in agreement with Ref.~\cite{savrasov_electron-phonon_1996,cazorla_melting_2008}. Instead, Ref.~\cite{zeng_lattice_2010} finds PBE phonon dispersions closer to experiment than ours.

\subsection{Thermal equation of state}

\begin{figure}[H]
\centering
\includegraphics[width=\linewidth]{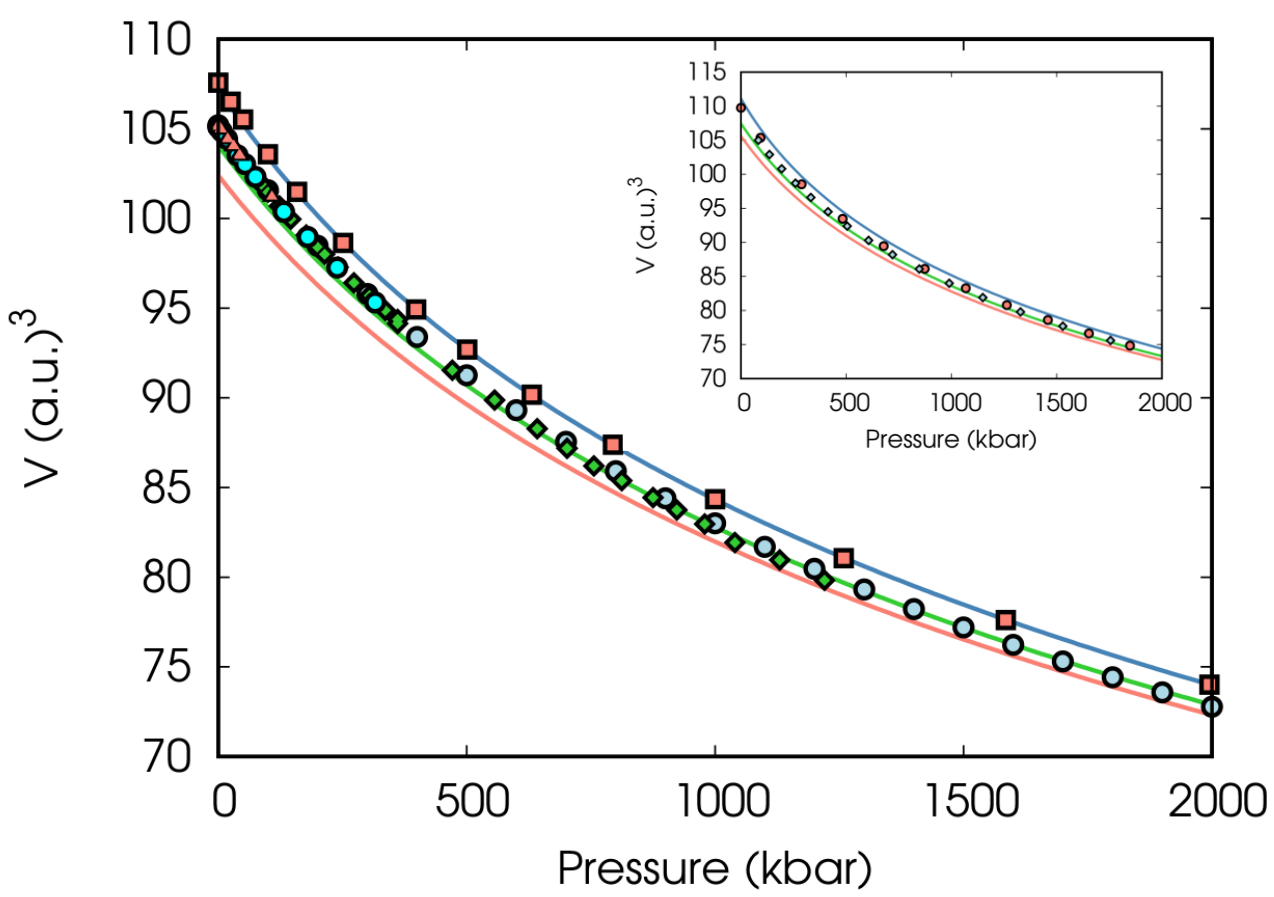}
\caption{$V$ as a function of $p$ at $300$ K obtained by LDA (red), PBEsol (green), and PBE (blue) compared with experiment (blue circles~\cite{hixson_shock_1992}, cyan circles,~\cite{litasov_thermal_2013_2}
red triangles,~\cite{zhao_thermoelastic_2000} and green diamonds~\cite{dewaele_compression_2008}) and previous calculations (red squares (PBE) from Ref.~\cite{wang_calculated_2000}). In the inset the same curves 
at $2000$ K compared with the predictions
of  Ref.~\cite{litasov_thermal_2013_2} (red circles)
and of Ref.~\cite{dorogokupets_near-absolute_2012} (blue diamonds).
}
\label{fig:eos_v}
\end{figure}
In Fig.~\ref{fig:eos_v}, we show the $300$ K equation of state.
Our PBE curve agrees with the linearized augmented plane wave (LAPW) all-electrons PBE calculation of Ref.~\cite{wang_calculated_2000} and with the PW91~\cite{perdew_accurate_1992} results of Ref.~\cite{zeng_lattice_2010}. The experimental results are all in good agreement among themselves with the shock wave data arriving beyond 2000 kbar. At each pressure, the PBE curve overestimates the volume, 
while LDA underestimate it as we have discussed for zero pressure. The PBEsol volume is still slightly underestimated at zero pressure, but the agreement 
with experiment increases with pressure. The same behaviour is found at $2000$ K as we show in the inset. Here we compare our equation
of state with the models of Ref.~\cite{sokolova_self_2013,dorogokupets_near-absolute_2012} and~\cite{litasov_thermal_2013_2}. The two models agree with each other and
with the ab-initio results especially using PBE and PBEsol. 
\begin{figure}[H]
\centering
\includegraphics[width=\linewidth]{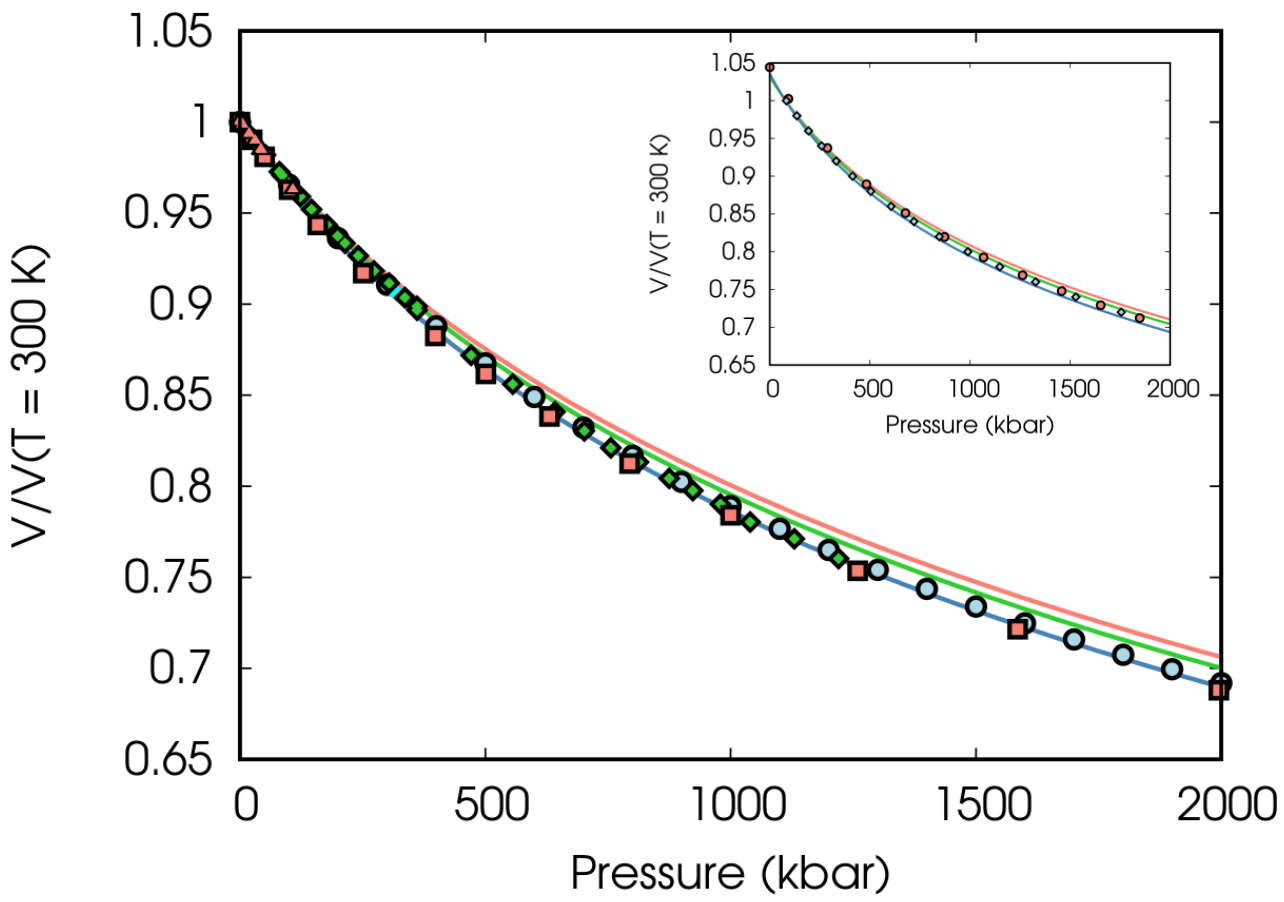}
\caption{$V /V_0$ as a function of $p$ at $300$ K obtained by LDA (red), PBEsol (green), and PBE (blue) compared with experiment (blue circles~\cite{hixson_shock_1992}, cyan circles,~\cite{litasov_thermal_2013_2}
red triangles,~\cite{zhao_thermoelastic_2000} and green diamonds~\cite{dewaele_compression_2008}) and previous PBE calculations (red~\cite{wang_calculated_2000} squares). In the inset the same curves 
at $2000$ K compared with the predictions
of Ref.~\cite{litasov_thermal_2013_2} (red circles)
and of Ref.~\cite{dorogokupets_near-absolute_2012} (blue diamonds).
}
\label{fig:eos}
\end{figure}

In Fig.~\ref{fig:eos} we present the equation of state removing the error due to the
equilibrium volume by plotting $V/V_0$ as a function of 
pressure (here $V_0$ is the equilbrium volume at $300$ K, close to that reported in Table~1 of the paper). As in the case of tungsten, here the picture
changes, with PBE that follows the experimental
data in all the pressure range. Both LDA and PBEsol
are in this case slightly above the experiment. In the inset we show the equation of state at
$2000$ K and compare with the model data presented in the inset of Fig.~\ref{fig:eos_v}. At this
temperature the model points are
equally close to PBE and PBEsol.
At zero pressure the theoretical data are slightly below the model point of Ref.~\cite{litasov_thermal_2013_2} reflecting the underestimation of the thermal expansion with respect to experiment (see below).

\subsection{Thermal expansion}

\begin{figure}[H]
\centering
\includegraphics[width=\linewidth]{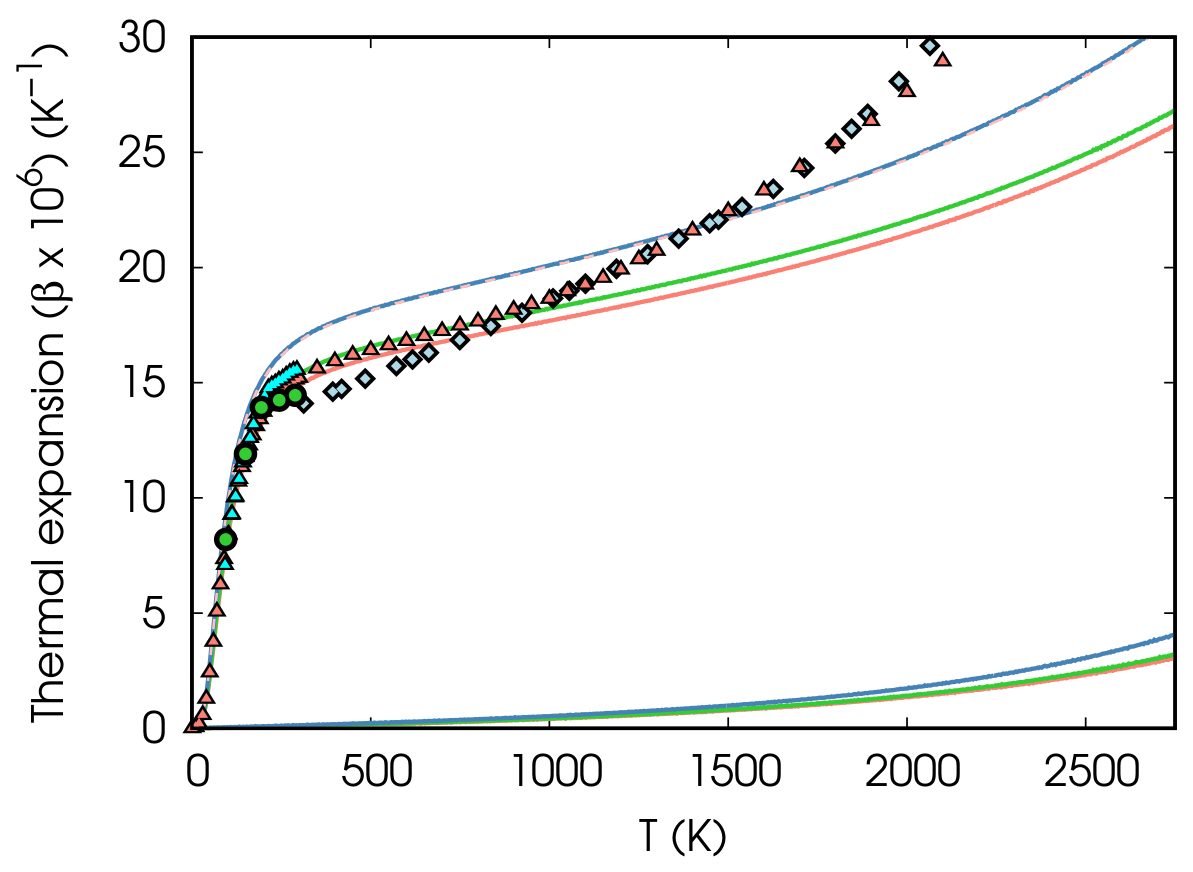}
\caption{Temperature dependent thermal expansion calculated by  
LDA (red curve), PBEsol (green curve), and PBE (blue curve) 
compared with the experimental data reported in Ref.~\cite{nix_thermal_1942} (cyan triangles) Ref.~\cite{kirby_american_1973} (green circles), Ref.~\cite{miiller_thermal_1985}
(blue diamonds), Ref.~\cite{bodryakov_correlation_2014} 
(red triangles). The thinner lines at the bottom indicate the differences
between the thermal expansion calculated including or neglecting the electronic excitations term in the free energy. Pink dashed line on the PBE curve shows the result obtained with $8$ geometries and $\Delta a=0.02$ a.u.}.
\label{fig:th_exp_t0}
\end{figure}
In Fig.~\ref{fig:th_exp_t0}, we compare with experiment the thermal expansion at zero pressure. We have reported only the commonly accepted data,
as those of Refs.~\cite{nix_thermal_1942,bodryakov_correlation_2014,kirby_american_1973,miiller_thermal_1985} and not the new measurements proposed by Ref.~\cite{zhao_thermoelastic_2000}.
Our LDA and PBEsol data match the accepted values up to $1000$ K while PBE is slightly higher. The PW91 data of Ref.~\cite{zeng_lattice_2010} (not reported here) are more close to our LDA data, than to the PBE data and match the experimental thermal expansion up to $1500$ K.
After these temperatures, anharmonic phonon-phonon interaction effects 
are expected to become important and therefore
our ab-initio data cannot follow experiment.
In the picture we show also the electronic
excitation contribution to the thermal 
expansion, calculated adding or neglecting
the electronic term in the free energy. As can be seen from the figure at $3000$ K it is of the
order of $5 10^{-6}$ (1/K), but at $1500$ K
its contribution is negligible.

\begin{figure}[H]
\centering
\includegraphics[width=\linewidth]{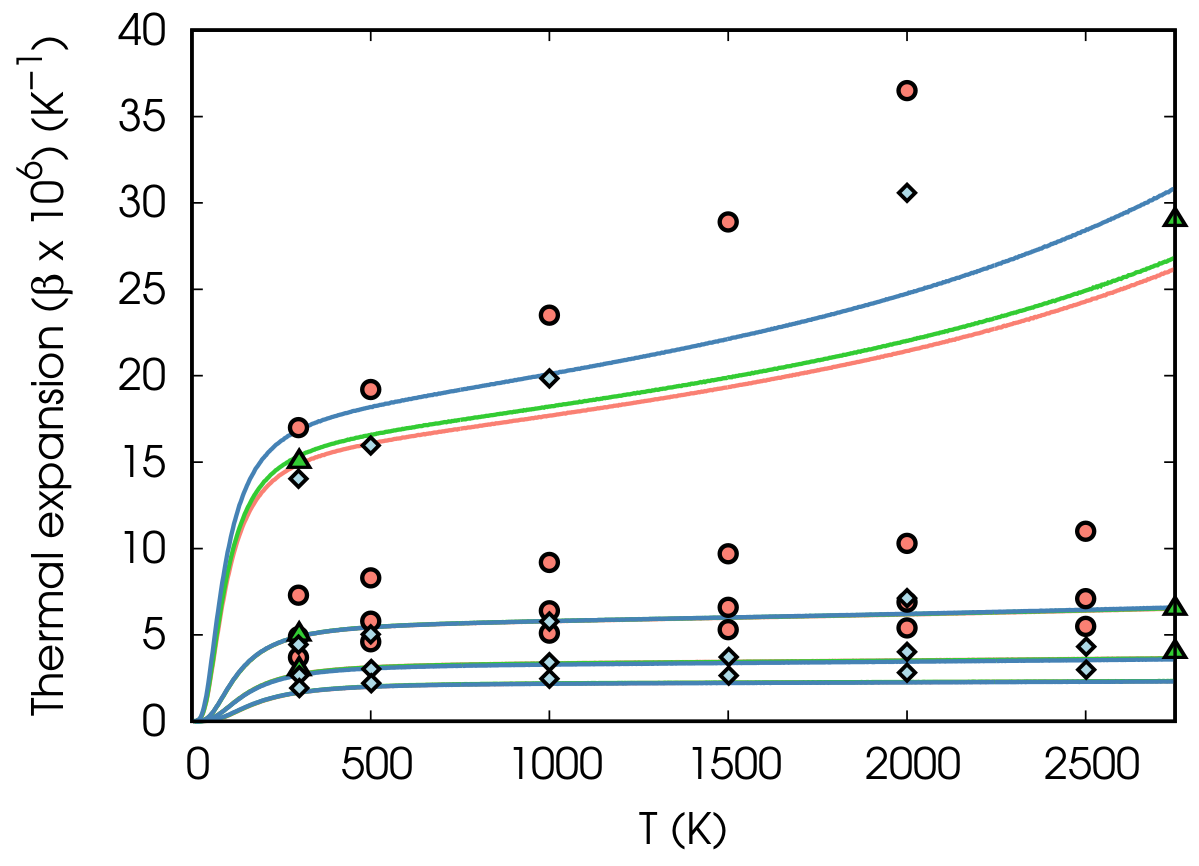}
\caption{Temperature dependent thermal expansion computed within the LDA (red line), PBEsol (green line), and PBE (blue line).
The different curves (from top to bottom) correspond to pressures from $0$ kbar to $3000$ kbar in steps $\Delta p=1000$ kbar. Theory is compared with the models of Ref.~\cite{litasov_thermal_2013_2} (red circles) and of Ref.~\cite{sokolova_self_2013} (blue diamonds) and with the ab-initio PBE calculations of Ref.~\cite{zeng_lattice_2010} (green triangles).
}
\label{fig:therm_exp}
\end{figure}

In Fig.~\ref{fig:therm_exp}, we show the thermal expansion at $0$ kbar, $1000$ kbar, $2000$ kbar,
and $3000$ kbar and compare our
data with the models of Ref.~\cite{litasov_thermal_2013_2} and
of Ref.~\cite{sokolova_self_2013}.
For the first we use the data reported in 
their Tab.~V, while for the latter we use
the data reported in Tab.~7 for $0$ kbar
and $1000$ kbar and data calculated by us
with the parameters reported at pag.193 
at $2000$ kbar and $3000$ kbar.
In the same figure we report also the PW91
values of Ref.~\cite{zeng_lattice_2010}.
We find that the three functionals differ 
appreciably at $0$ kbar, but the differences are smaller at higher pressures. The pressure dependence
of the thermal expansion agrees with the
PW91 findings of Ref.~\cite{zeng_lattice_2010}.
The points of the two models differ at high pressure, with the model of Ref.~\cite{litasov_thermal_2013_2} that decreases more slowly with pressure. Our data agree very well with the 
model of Ref.~\cite{sokolova_self_2013} until
about $1000$ K, but our curves are below the model points at higher temperatures. This seems
due to missing anharmonic phonon-phonon interaction terms in our calculation.

\subsection{Heat capacity}

\begin{figure}[H]
\centering
\includegraphics[width=\linewidth]{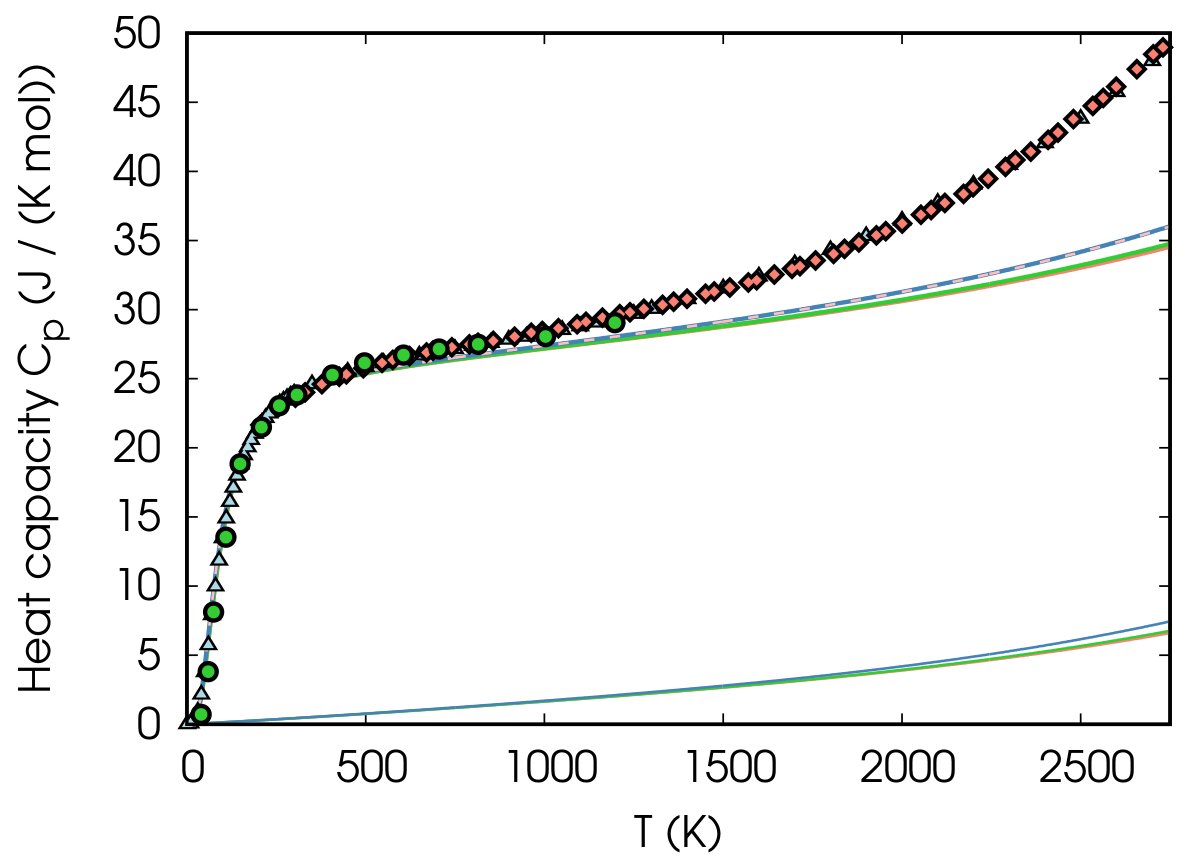}
\caption{LDA (red line), PBEsol (green line), and PBE (blue line)
temperature dependent isobaric heat capacity compared with
experiment from
Ref.~\cite{kirby_american_1973} (green circles), Ref.~\cite{guillermet_analysis_1991} (red diamonds) and Ref.~\cite{bodryakov_correlation_2014} (blue triangles).
Thin lines at the bottom (with the same color conventions) indicate the contribution
of electronic themal excitations to the heat capacity.
Pink dashed line on the PBE curve shows the result obtained with $8$ geometries and $\Delta a=0.02$ a.u.}.
\label{fig:cp_compare}
\end{figure}

In Fig.~\ref{fig:cp_compare}, we compare the isobaric heat capacity with experiment.
The three functionals gives very similar values of this quantity and theory and experiment are in good agreement until about $1000$ K.
As can be seen from the figure at $2500$ K 
our value of $33.2$ J/(K$\cdot$ mol) has a contribution of electronic excitation of about $5.7$ J/(K$\cdot$ mol), while the quasiharmonic term contributes
for other $2.5$ J/(K$\cdot$ mol), while the
experimental value is $44.1$ J/(K$\cdot$ mol) so that the anharmonic phonon-phonon interaction and possibly defects should give a contribution of about $10.9$
J/(K$\cdot$ mol).

\begin{figure}[H]
\centering
\includegraphics[width=\linewidth]{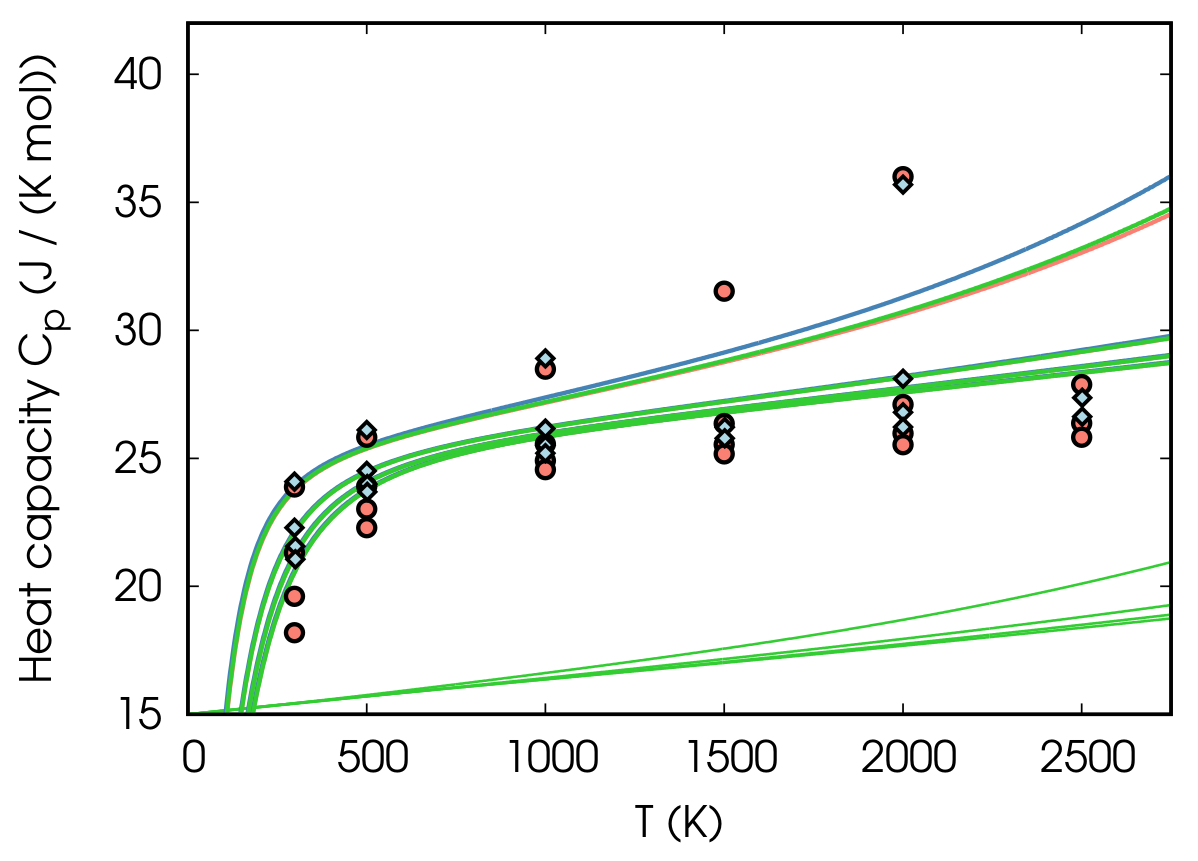}
\caption{LDA (red line), PBEsol (green line), 
and PBE (blue line) temperature dependent isobaric heat capacity.
From top to bottom the curves correspond to pressures from $0$ kbar to $3000$ kbar in steps $\Delta p=1000$ kbar. 
DFT calculations are compared with the model of Ref.~\cite{litasov_thermal_2013_2} (red circles) 
and of Ref.~\cite{dorogokupets_near-absolute_2012} (blue diamond).
The green lines at the bottom are the PBEsol electronic contribution to $C_V$ at the four pressures (from top to
bottom) shifted by $15$ J/(K $\cdot$ mol).}
\label{fig:cp} 
\end{figure}

In Fig.~\ref{fig:cp}, we report the isobaric heat capacity at several pressures. In the same figure we show also the electronic contribution
to which we added a fixed value of 
$15$ J/(K $\cdot$ mol) for picture clarity.
At zero pressure both models gives very
similar results which reproduce well experiment, but the behaviour with pressure is different, with the model of Ref.~\cite{litasov_thermal_2013_2} which
converges faster to the Dulong-Petit limit.
We find that the electronic contribution
decrease slowly with pressure and added to
the Dulong-Petit value gives a $C_p$ higher
than the point of Ref.~\cite{litasov_thermal_2013_2}.
It seems therefore that this model somewhat
underestimates the electronic contribution 
to the specific heat at high pressures.

\subsection{Bulk modulus}

\begin{figure}[H]
\centering
\includegraphics[width=\linewidth]{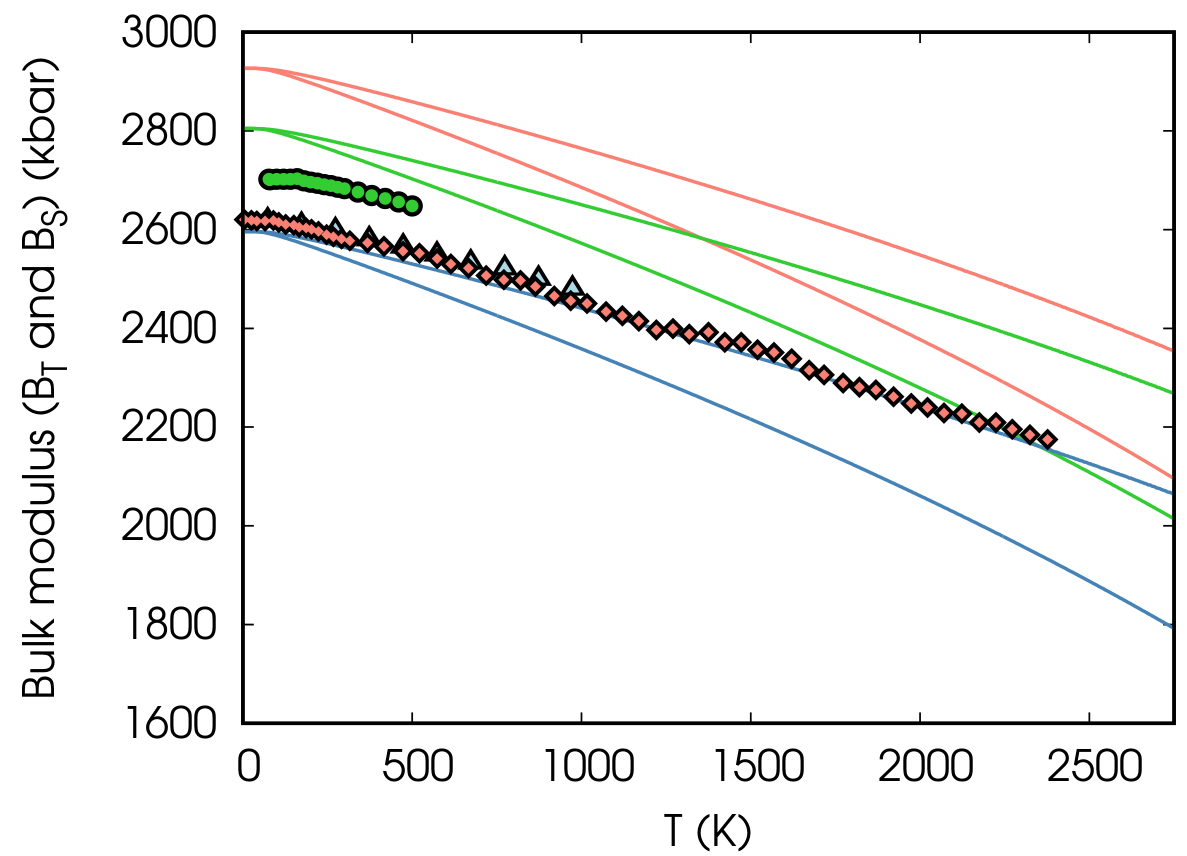}
\caption{Temperature dependent adiabatic and isothermal bulk moduli calculated within LDA (red line), PBEsol (green line), 
and PBE (blue line) compared with the experimental adiabatic data of Ref.~\cite{dickinson_temperature_1967} (blue triangles), Ref.~\cite{bolef_elastic_1962} (green circles) and Ref.~\cite{bujard_elastic_1981} (red diamonds). For each functional the higher curve is the adiabatic bulk modulus.}
\label{fig:bulk_modulus}
\end{figure}

\begin{figure}[H]
\centering
\includegraphics[width=\linewidth]{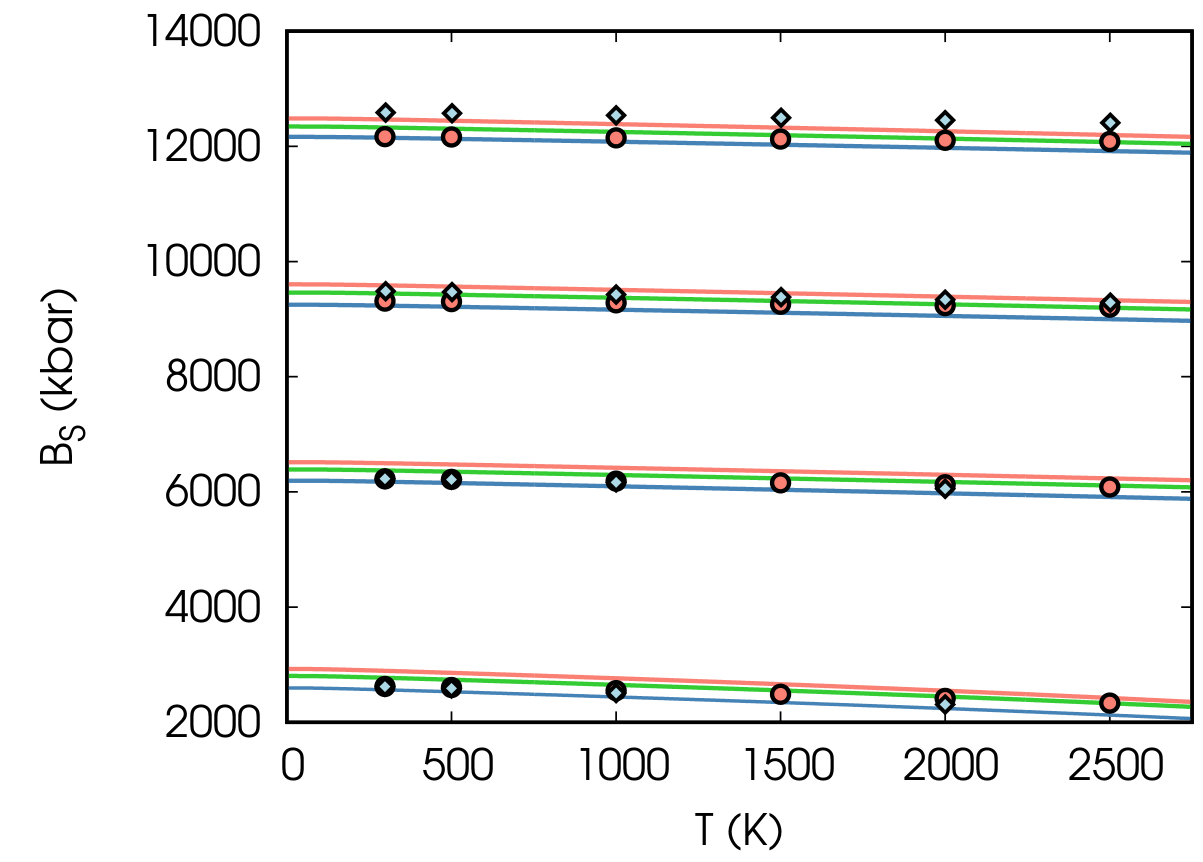}
\caption{Temperature dependent adiabatic bulk modulus calculated within the LDA (red lines), PBEsol (green lines), and PBE (blue lines). From bottom to top the curves correspond to pressures from $0$ kbar to $3000$ kbar in steps $\Delta p=1000$ kbar.
DFT calculations are compared with the model predictions of Ref.~\cite{litasov_thermal_2013_2} (red circles) and of Ref.~\cite{sokolova_self_2013} (blue diamond).
}
\label{fig:bulk_modulus_p}
\end{figure}

In Fig.~\ref{fig:bulk_modulus}, we compare the adiabatic bulk modulus with experiment. The isothermal bulk modulus is plotted for reference. The PBE bulk modulus is in very good agreement
with experiment and also the temperature dependence seems to be well accounted for
by theory. LDA and PBEsol instead overestimate
the bulk modulus. The decrease of the experimental bulk modulus from $25$ K to $2225$ $K$
is 410 kbar ($16$ \%), while the LDA, PBEsol, and PBE values are $432$ ($15$ \%), $407$ kbar ($15$ \%), and $405$ kbar ($16$ \%), respectively.
In Fig.~\ref{fig:bulk_modulus_p}, we report the adiabatic bulk modulus at several pressures. 
Our data are compared with the models of Ref.~\cite{litasov_thermal_2013_2} and
of Ref.~\cite{dorogokupets_near-absolute_2012} (only at $0$ kbar and $1000$ kbar).
In all the range of pressures the agreement between ab-initio values and model values is very good with PBE which remain always quite close to the model points and PBEsol and PBE slightly above.
For this quantity the differences between
functionals do not decrease with pressure.

\subsection{Gr\"uneisen parameter}
\begin{figure}[H]
\centering
\includegraphics[width=\linewidth]{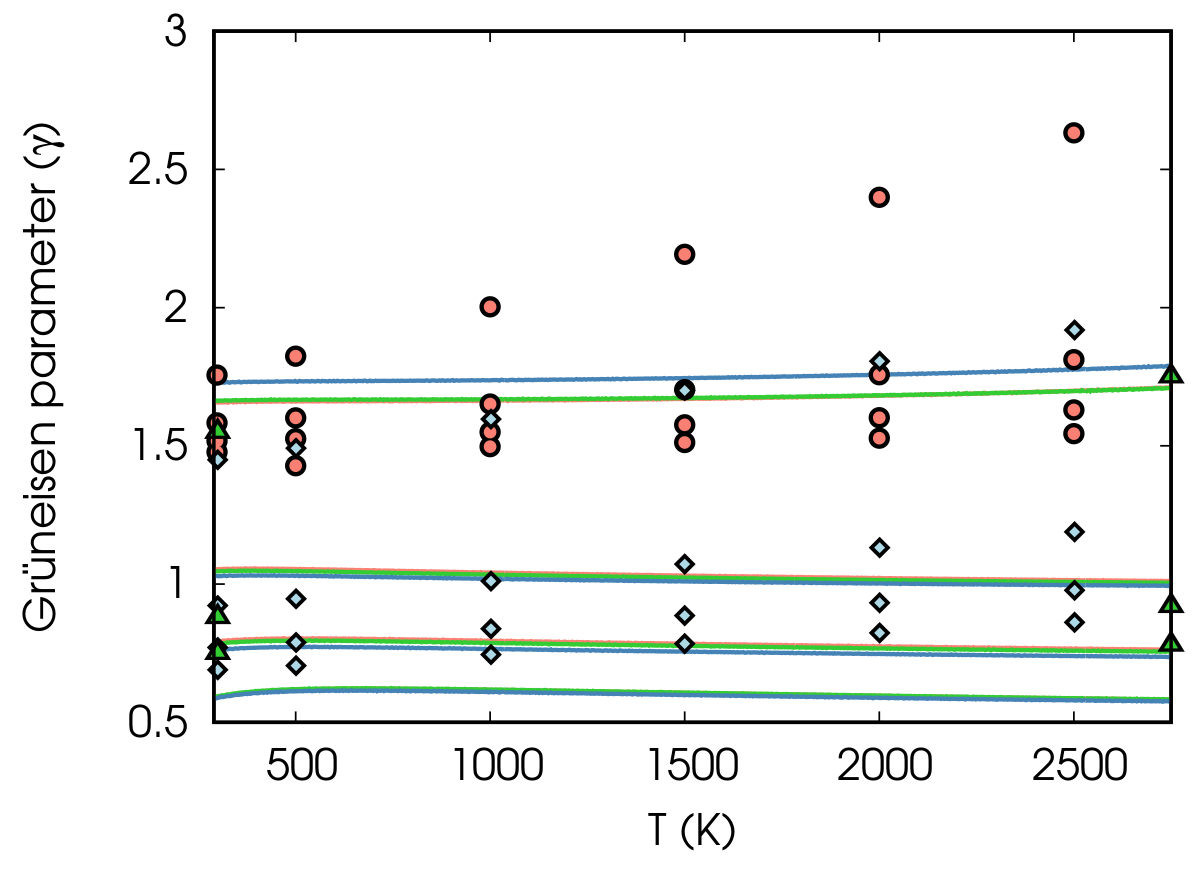}
\caption{Temperature dependent average Gr\"uneisen parameter calculated within the LDA (red line below the green line), PBEsol (green line) and PBE (blue line). From top to bottom the curves correspond to pressures
from $0$ kbar to $3000$ kbar in steps $\Delta p=1000$ kbar. DFT calculations are compared with the model predictions of Ref.~\cite{litasov_thermal_2013_2} (red circles) and of Ref.~\cite{dorogokupets_near-absolute_2012} (blue diamond) and with the ab-initio PBE calculations of Ref.~\cite{zeng_lattice_2010} (green triangles). 
}
\label{fig:average_grun}
\end{figure}

In Fig.~\ref{fig:average_grun}, we report
the average Gr\"uneisen parameter as 
a function of temperature and compare it
with previous ab-initio calculation~\cite{zeng_lattice_2010} 
an with the model data of Ref.~\cite{litasov_thermal_2013_2}
and of 
Ref.~\cite{dorogokupets_near-absolute_2012}
The agreement with the PW91 calculation
is quite good expecially at high pressure.
The major difference is at 0 kbar and
300 K where the value of Ref.~\cite{zeng_lattice_2010} is closer
to the model of Ref.~\cite{dorogokupets_near-absolute_2012} while our value is closer to the model
Ref.~\cite{litasov_thermal_2013_2}.
With temperature the model data and the ab-initio calculation increase
faster than our values and at 0 kbar
and 2750 K our value coincide with previous
ab-initio result, but remains lower
than both models.
The model of Ref.~\cite{litasov_thermal_2013_2} has average Gr\"uneisen parameters decreasing much slowly with pressure than both the ab-initio results and the model of Ref.~\cite{dorogokupets_near-absolute_2012}. The ab-initio results however seems to
remain more constant with temperature than the results of Ref.~\cite{dorogokupets_near-absolute_2012}. This in part could be due to the missing anharmonic phonon-phonon terms
in the ab-initio calculation.

\section{Numerical issues}

\begin{figure}
\begin{minipage}{0.999\textwidth} 
\includegraphics[width=.49\textwidth,keepaspectratio]{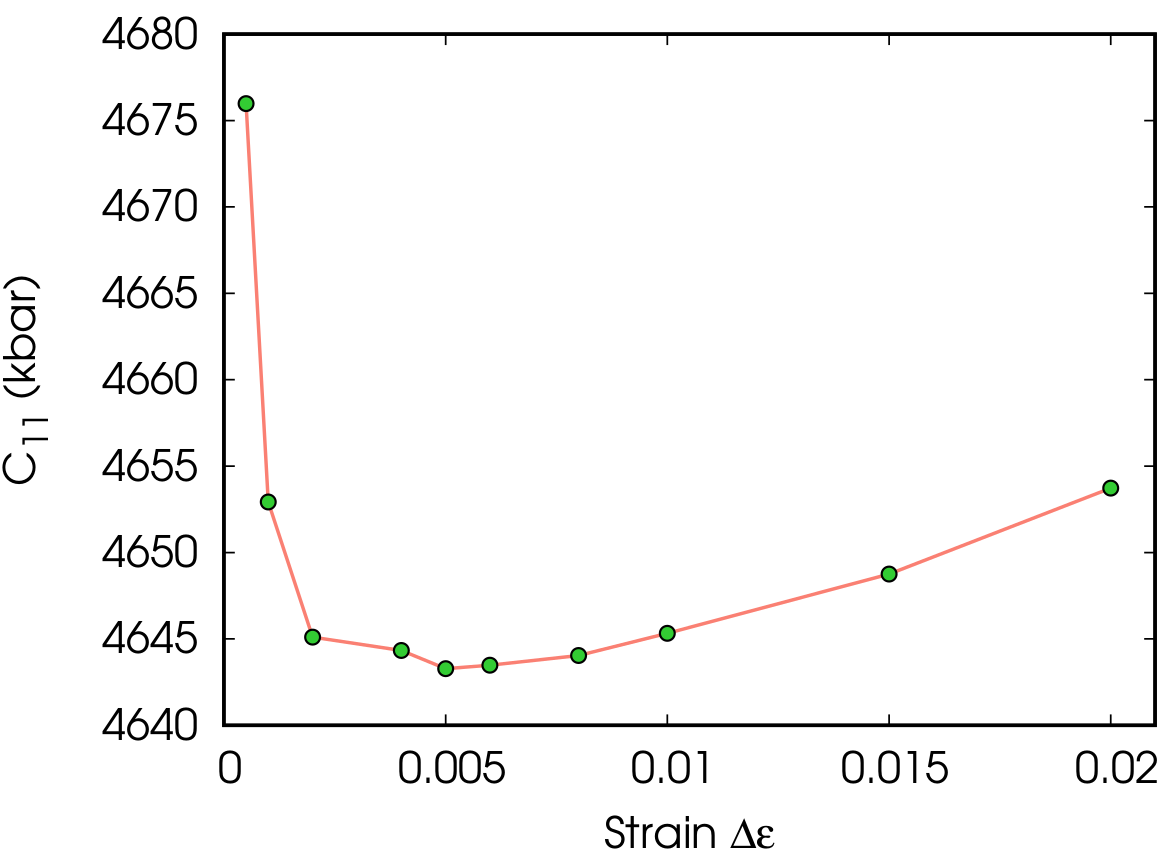}\hfill%
\includegraphics[width=.49\textwidth,keepaspectratio]{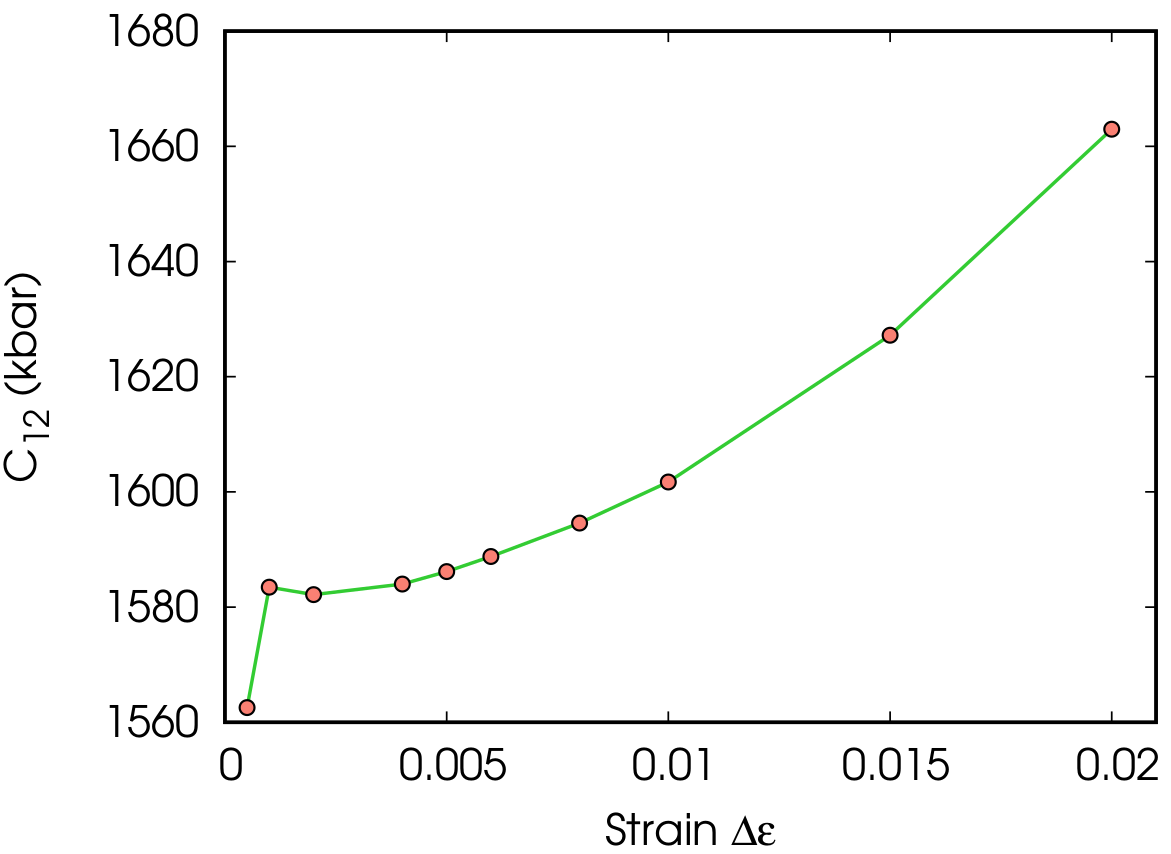}\\
\includegraphics[width=.49\textwidth,keepaspectratio]{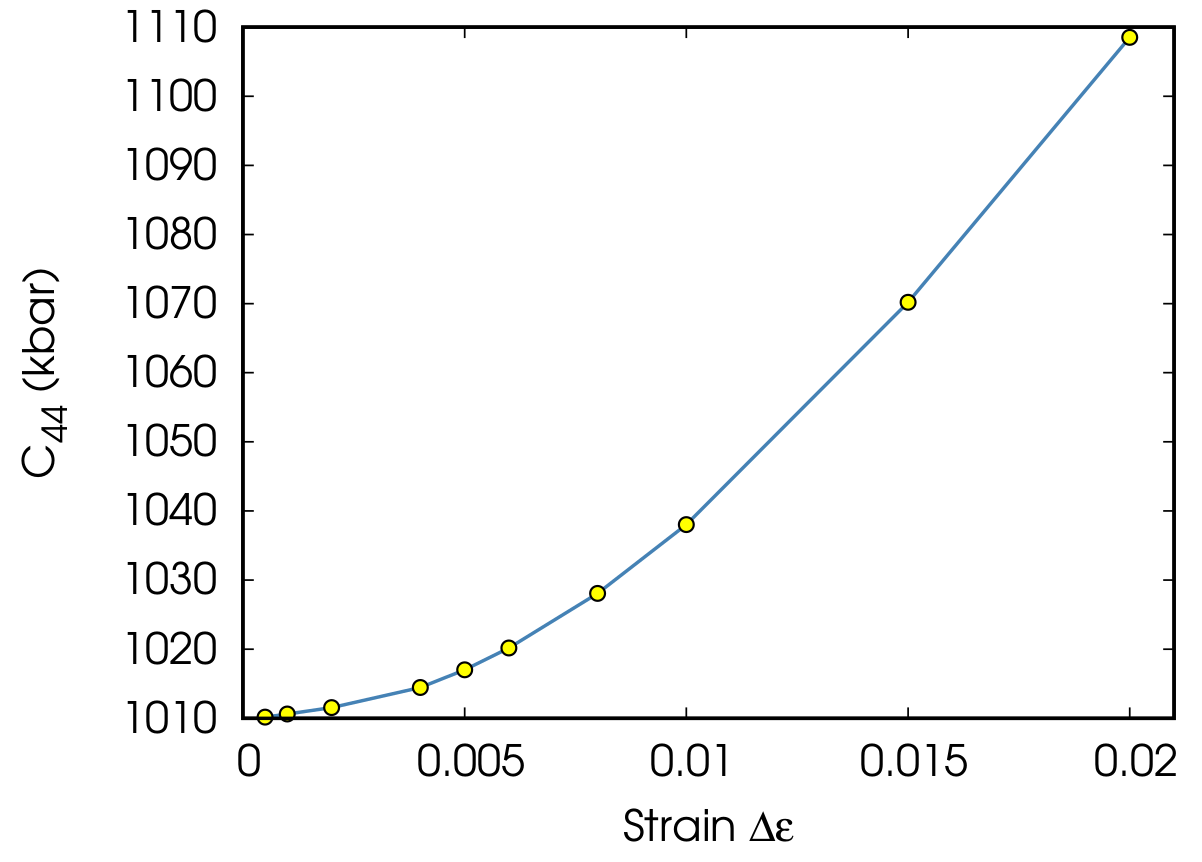}\hfill%
\end{minipage}
\caption{Value of the $T=0$ K elastic constants calculated with different values of the strain interval for each strain type.}
\label{teststrain}
\end{figure}

The choice of the interval between strains used
to sample the energy might affect the final value of the elastic constants. We show in Fig.~\ref{teststrain} the dependence of the
three elastic constants on this parameter.
As can be seen from the figure, when the strain interval is small enough to avoid the high order terms in the energy versus strain curves that appear at large $\Delta \varepsilon$, the values of the elastic constants converge to a constant value.
In principle, we should compute the limit 
for $\Delta \varepsilon \rightarrow 0$, but 
as can be seen in the curve for $C_{11}$, if the value
of $\Delta \varepsilon$ is too small, the fit of the
energy becomes also inaccurate. 
In the paper we used a value $\Delta\varepsilon=0.005$,
that gives values of elastic constants close to the limit of small $\Delta\varepsilon$ without introducing significant errors. We note that in these figures we are using an enlarged scale, but on the 
scale of elastic constants used in the paper, these curves appear almost flat.

We have also studied the dependence of our thermodynamic properties on the distance $\Delta a$ among the geometries used to compute phonons.
The $\Delta a$ used in the paper is $0.1$ a.u., but
also a $\Delta a=0.2$ a.u. gives very similar results as
we show for PBE in Fig.~\ref{fig:th_exp_t0} and in 
Fig.~\ref{fig:cp_compare}.

\bibliography{aipsamp}